\documentclass[11pt,a4paper]{article}
\usepackage{jheppub}
\usepackage{amsmath}
\usepackage{multicol,bbm}
\usepackage{enumerate}
\usepackage{bibentry}
\usepackage[toc,page]{appendix}
\usepackage{mathbbol}

\graphicspath{{./Figs/}}

\newcommand{\beqs}{\begin{equation*}}
\def\beq{\begin{equation}}

\newcommand{\eeqs}{\end{equation*}}
\def\eeq{\end{equation}}

\def\beq{\begin{equation}}
\def\eeq{\end{equation}}
\def\be{\begin{equation}}
\def\be{\begin{equation}}
\def\ee{\end{equation}}
\def\ba{\begin{eqnarray}}
\def\ea{\end{eqnarray}}
\def\bea{\begin{eqnarray}}
\def\eea{\end{eqnarray}}
\def\eq{\begin{equation}}
\def\eqe{\end{equation}}
\def\eqa{\begin{eqnarray}}
\def\eqae{\end{eqnarray}}
\def\la{\langle}
\def\ra{\rangle}
\def\beqa{\begin{eqnarray}}
\def\eeqa{\end{eqnarray}}
\def\ie{\begin{align}}
\def\fe{\end{align}}

\newcommand{\beqas}{\begin{eqnarray*}}

\newcommand{\eeqas}{\end{eqnarray*}}

\usepackage{subfigure}
\usepackage{tikz}
\usetikzlibrary{decorations.pathmorphing}
\usetikzlibrary{decorations.markings}
\usepackage{tikz}
\usetikzlibrary{arrows,shapes}
\usetikzlibrary{trees}
\usetikzlibrary{patterns}
\usetikzlibrary{matrix,arrows} 				
\usetikzlibrary{positioning}				
\usetikzlibrary{calc,through}				
\usetikzlibrary{decorations.pathreplacing}  
\usepackage{pgffor}	
\usetikzlibrary{decorations.pathmorphing}	
\usetikzlibrary{decorations.markings}
\tikzset{
    4Dline/.style={decorate, decoration={snake}, draw},
	provector/.style={decorate, decoration={snake,amplitude=2.5pt}, draw},
        smallvector/.style={decorate, decoration={snake,amplitude=1.5pt,post length=0.5mm}, draw},
    3Dline/.style={draw=black, postaction={decorate},
        decoration={markings,mark=at position .55 with {\arrow[draw=black]{>}}}},
    3Dlinenoarrow/.style={draw=black},
    scalar/.style={dashed,draw=black, postaction={decorate},
        decoration={markings,mark=at position .55 with {\arrow[draw=black]{>}}}},
}
\tikzstyle{block} = [draw, rectangle,  minimum height=3em, minimum width=6earticlem]
\graphicspath{{./Figs/}} 
\usepackage{tikz}
\usetikzlibrary{decorations.pathreplacing,decorations.markings}
\tikzset{
  on each segment/.style={
    decorate,
    decoration={
      show path construction,
      moveto code={},
      lineto code={
        \path [#1]
        (\tikzinputsegmentfirst) -- (\tikzinputsegmentlast);
      },
      curveto code={
        \path [#1] (\tikzinputsegmentfirst)
        .. controls
        (\tikzinputsegmentsupporta) and (\tikzinputsegmentsupportb)
        ..
        (\tikzinputsegmentlast);
      },
      closepath code={
        \path [#1]
        (\tikzinputsegmentfirst) -- (\tikzinputsegmentlast);
      },
    },
  },
  mid arrow/.style={postaction={decorate,decoration={
        markings,
        mark=at position .55 with {\arrow[#1]{stealth}}
      }}},
}

\title{The orthogonal momentum amplituhedron and ABJM amplitudes}

\author[a,b]{Yu-tin Huang,}
\author[a]{Ryota Kojima,}
\author[c]{Congkao Wen,}
\author[c]{Shun-Qing Zhang}

\affiliation[a]{Department of Physics and Center for Theoretical Physics, National Taiwan University, Taipei 10617, Taiwan}
\affiliation[b]{Physics Division, National Center for Theoretical Sciences, Taipei 10617, Taiwan}
\affiliation[c]{Centre for Theoretical Physics, Department of Physics and Astronomy, Queen Mary University of London,
London, E1 4NS, UK}

\emailAdd{yutinyt@gmail.com}
\emailAdd{ryotakojima914@gmail.com }
\emailAdd{c.wen@qmul.ac.uk}
\emailAdd{shun-qing.zhang@qmul.ac.uk}

\abstract{In this paper, we introduce the momentum space amplituhedron for tree-level scattering amplitudes of ABJM theory. We demonstrate that the scattering amplitude can be identified as the canonical form on the space given by the product of positive orthogonal Grassmannian and the moment curve. The co-dimension one boundaries of this space are simply the odd-particle planar Mandelstam variables, while the even-particle counterparts are ``hidden" as higher co-dimension boundaries. Remarkably, this space can be equally defined through a series of ``sign flip" requirements of the projected external data, identical to ``half" of four-dimensional $\mathcal{N}=4$ super Yang-Mills theory (sYM). Thus in a precise sense the geometry for ABJM lives on the boundary of $\mathcal{N}=4$ sYM.  We verify this relation through eight-points by showing that the BCFW triangulation of the amplitude tiles the amplituhedron. The canonical form is naturally derived using the Grassmannian formula for the amplitude in the $\mathcal{N}=4$ formalism for ABJM theory. }
\begin{document}
\begin{flushright}
\hfill{QMUL-PH-21-48} \end{flushright}
\maketitle 

\section{Introduction}
Scattering amplitudes in $\mathcal{N}=6$ Chern-Simons matter theory (often termed ABJM)~\cite{Aharony:2008ug, Hosomichi:2008jb}, have long been an interesting close cousin of those in four-dimensional $\mathcal{N}=4$ super Yang-Mills (sYM), mimicking its hidden structures with modifications tailored to the unique features of three-dimensional kinematics. For example the all multiplicity tree-amplitude worldsheet formula of Witten-RSV~\cite{Witten:2003nn, Roiban:2004yf}, has a mirror image in ABJM theory~\cite{Huang:2012vt}. Similarly the SU(4$|$4) dual-superconformal (and its full Yangian embedding) invariance of tree-level amplitude and loop-level integrand of $\mathcal{N}=4$ sYM~\cite{Drummond:2008vq, Drummond:2009fd}, have their counterpart, the OSp(6$|$4) of the $\mathcal{N}=6$~\cite{Bargheer:2010hn, Huang:2010qy, Gang:2010gy}. As a consequence, the Grassmannian geometry that yields individual Yangian blocks~\cite{Arkani-Hamed:2009ljj}, once constrained to its orthogonal subspace yields the leading singularities of ABJM theory~\cite{Lee:2010du}. The stratification of the geometry admits a trivalent bi-partite graphical representation for the individual cells~\cite{Arkani-Hamed:2012zlh}, can also be applied to ABJM theory with the simplification of using medial graphs with quadratic vertices~\cite{Huang:2013owa, Huang:2014xza}.

As is apparent in the above, this hand in hand development appears to have as its boundary the extension to momentum twistor~\cite{Hodges:2009hk}. Indeed the latter was instrumental in the realization of \textit{amplituhedron}~\cite{Arkani-Hamed:2013jha, Arkani-Hamed:2013kca}, where the amplitude is identified as the canonical form on a positive geometry whose boundaries are given in momentum twistors. The difficulty lies in the nature of dual superconformal symmetry in three dimensions, which requires in addition to the introduction of dual variables for the conformal group Sp(4), but also the R-symmetry SO(6). This will appear to require a new set of twistor variables that do not have a kinematic origin. 

An alternative amplituhedron definition for tree-level amplitudes of $\mathcal{N}=4$ sYM was proposed directly in the spinor helicity kinematic space~\cite{Damgaard:2019ztj}, motivated by~\cite{He:2018okq}. This opens the possibility for the existence of a tree-level amplituhedron for ABJM theory directly in the three-dimensional kinematic space. In this paper we will present precisely such a geometric object, which we call as the {\it orthogonal momentum amplituhedron}. Consider the image of the following map:
\eq\label{YMap}
Y_{a}^A=\sum_{i=1}^n\; c_{ai}\Lambda_i^A\,, 
\eqe
where $a=1,\cdots,k$, $A=1,\cdots,k{+}2$ and $k=\frac{n}{2}$ ($n$ is even here). Here $Y_{a}^A$ lives in a subspace of the Grassmannian $G(k,k{+}2)$, that is the image of positive orthogonal Grassmannian $OG_{+}(k,2k)$ ($c_{ai}$ is an element of $OG_{+}(k,2k)$) mapped through the bosonic twistor variables $\Lambda_i^A$ living on a moment curve. As we will show, the boundary of this space is given by odd-particle planar Mandelstam variables $S_{i, i{+}1, i{+}2,\cdots, i{+} p}$ (for odd number of particles, $p$ is even), where 
\eq
S_{i,i{+}1,\cdots, i{+} p}=\sum_{i\leq j<l\leq i{+}p} (-1)^{j{+}l{+}1}\langle Y j l\rangle^2,\quad  \langle Y j l\rangle\equiv \epsilon_{A_{1}A_{2}\cdots A_{2{+}k}}Y_1^{A_1}Y_2^{A_2}{\cdots} Y_{k}^{A_{k}}\Lambda_j^{A_{1{+}k }}\Lambda_l^{A_{2{+}k }}\,.
\eqe
Note that while all planar Mandelstam variables are non-negative, only the vanishing of each odd-particle Mandelstam variable is co-dimension one boundary. The vanishing of even-particle Mandelstam variables is higher co-dimensional. This reflects the fact that the non-vanishing amplitudes in ABJM theory have an even number of particles, therefore the amplitudes only have factorization poles of odd-particle Mandelstam variables. The space is $(n{-}3)$-dimensional, as $Y^A_a$ given \eqref{YMap} satisfy the following conditions:
\eq
\sum_{i=1}^n \; (-1)^i (Y^\perp\cdot \Lambda^T)_i^\alpha (Y^\perp\cdot \Lambda^T)_i^\beta=0\,. 
\eqe

The amplitude is then identified with the volume function $\Omega_{2k,k}$ defined through 
\eq \label{eq:rel}
{\bf{\Omega}}^{3d}_{2k,k}\wedge d^3P\; \delta^3(P) =  \Omega_{2k,k} \left(\prod_{a=1}^{k} \langle Y_1Y_2\cdots Y_{k} d^2 Y_a\rangle \right)\delta^3(P)\, ,
\eqe
where ${\bf{\Omega}}^{3d}_{2k,k}$ is the $(n{-}3)$-dimensional canonical form on $OG_{+}(k,2k)$ whose co-dimension one boundaries, via the map in (\ref{YMap}) are the planar odd-particle Mandelstam variables. The subspace defined through the map in (\ref{YMap}), can be carved out directly in $Y$ space via the non-negativity of $\langle Y i i{+}1\rangle$, and a series of sign pattern as well as the momentum conservation:
\eqa
\{\langle Y12\rangle, \langle Y13\rangle, \cdots,\langle Y1n\rangle\},\quad {\rm having}\;k\; {\rm sign}\;{\rm flips} \, , \nonumber\\
\sum_{i=1}^n\;(-1)^i\langle Yia\rangle\langle Yib\rangle=0,\quad{\rm for}\; a,b=1,\cdots,n \, .
\eqae
Note that this is identical to the sign flipping conditions associated with half of the momentum amplituthedron of $\mathcal{N}=4$ sYM~\cite{Damgaard:2019ztj}. In particular, the amplituhedron geometry for the  four-dimensional theory is given by $(Y,\tilde{Y})\in (Gr(n{-}k, n{-}k{+}2), Gr(k,k{+}2))$, with $Y, \tilde{Y}$ satisfying $k{-}2$ and $k$ sign-flip patterns respectively. Thus with $k=\frac{n}{2}$, we see that the orthogonal amplituhedron geometry for the ABJM theory can be identified with $\tilde{Y}$, with the additional constraint associated from momentum conservation. Thus \textit{the orthogonal momentum amplituhedron is simply a kinematic projection of the momentum amplituhedron geometry for four-dimensional theories}.

We verify the above proposal through the BCFW construction, which identifies the tree-level amplitude as a particular combination of cells of the orthogonal Grassmannian~\cite{Huang:2013owa, Huang:2014xza}. We first confirm that the BCFW cells via the map in  (\ref{YMap}) tile the whole space. This is checked numerically at eight points, where each point in the image for the top cell lies only in one of the BCFW cells, vice versa. Next we identify the canonical form as
\eq\label{Main}
{\bf{\Omega}}^{3d}_{2k,k}=\sum_\sigma \int_{\mathcal{C}_\sigma} {d^{k \times 2k }C \over {\rm Vol}(GL(k))}\; \frac{M_{1,3,5,\cdots,n-1}}{M_1M_2\cdots M_{k}}\delta^{\frac{k(k+1)}{2}}(C^TC) \bigg\rvert_{Y= c\cdot \Lambda}\,,
\eqe
where the contour $\mathcal{C}_\sigma$ localizes on the various BCFW cells labelled by $\sigma$, and $M_{1,3,5,\cdots,n-1}$ is a $k\times k$ involving columns $\{1,3,5,\cdots,n-1\}$, and similarly $M_i$ is the minor  involving consecutive columns $\{i, i{+}1,\cdots, i{+}k{-}1\}$. Note that the integrand is simply the original orthogonal integral introduced in~\cite{Lee:2010du}, but reduced to $\mathcal{N}=4$  SUSY, which leads to the numerator $M_{1,3,5,\cdots,n-1}$. The union of these forms then gives the BCFW triangulation of the amplituhedron. Since the BCFW cells tile the space for $Y$, with the contour $\mathcal{C}$ encircling these cells,  (\ref{Main}) gives the correct  canonical form that can be lifted to the volume form for the amplituhedron via the relation \eqref{eq:rel}. It is intriguing that the canonical form on $OG_+(k,2k)$ is more naturally derived using the $\mathcal{N}=4$ formalism. Note that this is natural from the viewpoint of exchanging $\eta \rightarrow d\lambda $, similar to~\cite{He:2018okq}. Indeed the $n$-point amplitude is degree $n$ in $\eta$, and thus produces the $n$-form which can be matched to the volume form in kinematic space.

The rest of the paper is organised as follows. In the next section, we will briefly review basic properties of the momentum amplituhedron for four-dimensional  $\mathcal{N}=4$ sYM. In Section \ref{sec:ABJM-hedron}, we present the construction of  the orthogonal momentum amplituhedron geometry and its definition through the sign flipping.  In section \ref{sec:forms}, we discuss in detail canonical forms of  the  orthogonal momentum amplituhedron, and their associated singularities and the boundary structures of the amplituhedron geometry. In section \ref{sec:con}, we conclude and remark on future research directions.

\section{Review of the momentum amplituhedron for $\mathcal{N}=4$ sYM}
In this section, we will review the construction of the momentum amplituhedron for four-dimensional $\mathcal{N}=4$ sYM~\cite{Damgaard:2019ztj}.\footnote{For further study on the momentum amplituhedron, see \cite{Ferro:2020lgp, Damgaard:2020eox, Damgaard:2021qbi}.} The momentum amplituhedron $\mathcal{M}_{n,k}$ is defined as the image of the positive Grassmannian $G_+(k,n)$ through the map depending on the positive kinematics. Here, positive kinematics is defined as two sets of moment curves on which the external data lives: 
\begin{align}
(\Lambda^{\perp})^{\bar{A}}_i = i^{\bar{A}-1},\quad \tilde{\Lambda}^{\dot{A}}_i = i^{\dot{A}-1}\,.
\end{align}
 From this definition, we can see easily that all ordered minors of matrices $\tilde{\Lambda}, \Lambda^\perp$ are positive. When we extract the amplitudes, these matrices are identified as the bosonized kinematics:
\begin{align}
&\Lambda^A_i = \begin{pmatrix}
\lambda_i^{\alpha} \\
\phi^{1}_I \cdot \eta^I_i\\
\vdots\\
\phi^{n-k}_I \cdot \eta^I_i
\end{pmatrix}, \ \ \ A=1,\dots,n-k+2,\nonumber\\
&\tilde{\Lambda}^{\dot{A}}_i = \begin{pmatrix}
\tilde{\lambda}_i^{\dot{\alpha}} \\
\tilde{\phi}^{1}_{\hat{I}} \cdot \tilde{\eta}^{\hat{I}}_i \\
\vdots\\
\tilde{\phi}^{k}_{\hat{I}} \cdot \tilde{\eta}^{\hat{I}}_i
\end{pmatrix}, \ \ \ \dot{A}=1,\dots,k+2,
\end{align}
where $\eta,\tilde{\eta},\phi,\tilde{\phi}$ are Grassmann-odd variables. Here we use the non-chiral SUSY for describing $\mathcal{N}=4$ sYM superamplitudes, with a subgroup of R-symmetry $SU(2) \times SU(2)$ being manifest. Therefore $I=1,2, \quad \hat{I}=\hat{1}, \hat{2}$, and one may identify the R-symmetry index $I$ with the little group index $\alpha$ (and $\hat I$ with $\dot \alpha$), and superamplitudes become differential forms after the identification $\eta \rightarrow d\lambda, \tilde{\eta} \rightarrow d \tilde{\lambda}$ \cite{He:2018okq}. 

The momentum amplituhedron $\mathcal{M}_{n,k}$ is defined as a pair of Grassmannian elements $(\tilde{Y},Y)\in G(k,k+2)\times G(n-k,n-k+2)$:
\begin{align} \label{eq:Ymap_4d}
\tilde{Y}^{\dot{A}}_{\dot{a}} = \sum_{i=1}^n  c_{\dot{a} i} \tilde{\Lambda}^{\dot{A}}_{i}\, \qquad Y^A_a = \sum_{i=1}^n  c_{a i}^\perp \Lambda^A_i ,
\end{align} 
where $c_{\dot{a}i}$ are the elements of the positive Grassmannian $G_+(k,n)$ and $c^\perp_{a i}$ are the element of its orthogonal complement. Although the dimension of the $(\tilde{Y},Y)$ space is
\begin{align}
\text{dim}(G(k,k+2))+\text{dim}(G(n-k,n-k+2)) = 2(n-k)+2k=2n,
\end{align} 
the momentum amplituhedron $(\tilde{Y},Y)$ is satisfying the following relation:
\begin{align} \label{eq:momentumcons}
P^{\alpha \dot{\alpha}}=\sum_{i=1}^n\left(Y^\perp \cdot \Lambda^T\right)^{\alpha}_i\left(\tilde{Y}^\perp \cdot \tilde{\Lambda}^T\right)^{\dot{\alpha}}_i=0.
\end{align} 
Then the momentum amplituhedron has dimension $2n-4$.

The definition of the momentum amplituhedron implies particular sign patterns for $Y$ and $\tilde{Y}$ brackets 
\begin{align} \label{eq:signflips}
&\{\langle Y 12 \rangle,\langle Y 13 \rangle,\dots,\langle Y 1n \rangle\}\ \text{has $k-2$ sign flips,}\\
&\{[\tilde{Y}12],[\tilde{Y}13],\dots,[\tilde{Y}1n]\}\ \text{has $k$ sign flips.}
\end{align} 
Here we introduce the brackets 
\begin{align} 
&\langle Y ij \rangle=\epsilon_{A_1A_2\dots A_{n-k+2}}Y_{i_1}^{A_1}Y_{i_2}^{A_2}\dots Y_{i_{n-k}}^{A_{n-k}} \Lambda_i^{A_{n-k+1}}\Lambda_j^{A_{n-k+2}}, \nonumber\\
&[\tilde{Y}ij]=\epsilon_{\dot{A}_1\dot{A}_2\dots \dot{A}_{k+2}}\tilde{Y}_{i_1}^{\dot{A}_1}\tilde{Y}_{i_2}^{\dot{A}_2}\dots \tilde{Y}_{i_{k}}^{\dot{A}_{k}}\tilde{\Lambda}_{i}^{\dot{A}_{k+1}}\tilde{\Lambda}_{j}^{\dot{A}_{k+2}}\,.
\end{align} 
The co-dimension one boundaries are then simply 
\eq
\langle Y i i+1\rangle=[\tilde{Y} i i{+}1]=0,\quad{\rm and}\quad S_{i,i{+}1,\cdots,i{+}p}\equiv \sum_{i\leq j_1<j_2\leq i{+}p}\;\langle Y j_1 j_2\rangle[\tilde{Y} j_1j_2]=0 \, .
\eqe 
Note that as discussed in~\cite{Damgaard:2019ztj} it is crucial for the external kinematics to be ordered on the moment curve for the planar Mandelstams to be positive, and hence its zero being the boundaries.

In order to obtain scattering amplitudes from the momentum amplituhedron, we need to construct the canonical form $\bold{\Omega}_{n,k}$ with logarithmic singularities on all boundaries. The momentum amplituhedron $\mathcal{M}_{n,k}$ is $2n{-}4$ dimensional and therefore its canonical form $\bold{\Omega}_{n,k}$ has also the same degree. One then constructs the volume form 
\eq
V_{vol}=\prod_{a=1}^{n{-}k}\langle Y_1\dots Y_{n{-}k}d^2Y_a \rangle\prod_{\dot{a}=1}^k[\tilde{Y}_1\dots \tilde{Y}_kd^2\tilde{Y}_{\dot{a}} ]\Omega_{n,k}\,,
\eqe
through the relation 
\begin{align}\label{Relate} 
\bold{\Omega}_{n,k}\wedge d^4P \delta^4(P)=\prod_{a=1}^{n-k}\langle Y_1\dots Y_{n-k}d^2Y_a \rangle\prod_{\dot{a}=1}^k[\tilde{Y}_1\dots \tilde{Y}_kd^2\tilde{Y}_{\dot{a}} ]\Omega_{n,k}\delta^4(P)\,.
\end{align} 
The amplitude is obtained from the volume function $\Omega_{n,k}$, where we localize $(Y, \tilde{Y})$ to $(Y^*,\tilde{Y}^*)$:
\begin{align}
Y^*=\begin{pmatrix}
\mathbb{0}_{2\times(n-k)} \\
\mathbb{1}_{(n-k)\times(n-k)}
\end{pmatrix}, \ \ \ 
\tilde{Y}^*=\begin{pmatrix}
\mathbb{0}_{2\times(k)} \\
\mathbb{1}_{(k)\times(k)}
\end{pmatrix}. 
\end{align}
The amplitude can be obtained by integrating out the auxiliary fermionic variables $\phi$ and $\tilde{\phi}$ that we have introduced 
\begin{align}
\mathcal{A}_{n,k}^{\text{tree}}=\delta^4(p)\int d^2\phi^1 \dots d^2\phi^{n-k} \int d^2\tilde{\phi}^1 \dots d^2\tilde{\phi}^k  \, \Omega_{n,k}(Y^*,\tilde{Y}^*,\Lambda,\tilde{\Lambda})\,.
\end{align}

In practice one can use the BCFW triangulation to construct the form for $\bold{\Omega}_{n,k}$. We write 
\eq
\bold{\Omega}_{n,k}=\sum_\sigma \;\int_{\mathcal{C}_{\{M_\sigma\}}} \frac{d^{n\times k}c}{{\rm Vol}(GL(k))}{1\over M_{1}M_2\cdots M_n} \bigg\rvert_{Y= c^{\perp}\cdot \Lambda,\; \tilde{Y}=c\cdot \tilde{\Lambda}}\, ,
\eqe
where $\sigma$ labels the set of BCFW cells that constitute the tree amplitude, with each cell characterized by a set of vanishing minors $\{M_\sigma\}$ and hence the integration contour $\mathcal{C}_{\{M_\sigma\}}$. To obtain $\Omega_{n,k}$, one starts with the $G_+(k,n)$ top cell $c$, and solve for the set of vanishing minors associated with each cell $\{M_\sigma\}$. Next, momentum conservation in  (\ref{eq:momentumcons}) is used to constrain the top cell $c'$ of $G_{+}(n{-}k,k)$. We partially solve it so that $c'=c^\perp+\Delta$ where $\Delta$ would contain four unfixed parameters, which will be set to zero on the support of $\delta^4(P)$. Matching both sides of  (\ref{Relate}) allows us to fix $\Omega_{n,k}$.

\section{The orthogonal momentum amplituhedron}
\label{sec:ABJM-hedron}

In this section, we will introduce the orthogonal momentum amplituhedron geometry. We will define it in two ways. In the first way, we utilize the positive orthogonal Grassmannian $OG_+$ through the definition of $Y = C\cdot \Lambda$, with $C \in OG_+$. We will  also define the geometry by understanding its sign flipping structures.  The  canonical forms of the  orthogonal momentum amplituhedron and their relations to the amplitudes in the ABJM theory will be studied in the next section \ref{sec:forms}. 


\subsection{Definition of the orthogonal momentum amplituhedron}

In this section, we define the Orthogonal momentum amplituhedron. Again, we first consider the positive external data, where $\Lambda_i^A$ are $2k$ ordered points on an $k{+}2$-dimensional moment curve:  
\begin{equation}
\Lambda_i^A=x_i^{A-1}\,,
\end{equation}
where $x_i$s are arbitrary ordered points $x_1<x_2<\cdots<x_n$. This arrangement will be necessary for the planar Mandelstams to be positive as we will soon see.  For ABJM we will always have $2k\!=\!n$, thus from the get go the geometry is closely related to the middle sector (split helicity) of $\mathcal{N}=4$ sYM. 
As a result the moment matrix $\Lambda_i^A \in G(k{+}2,2k)$ will have all ordered minors being positive. This matrix will be identified as the bosonized kinematic variables
\begin{align}
\Lambda^A_i = \begin{pmatrix}
\lambda_i^{\alpha} \\
\phi^{a}_I \cdot \eta^I_i
\end{pmatrix},\quad A= (\alpha, a)=1,2, \ldots, k{+}2,
\end{align}
where we introduced $k$ auxiliary Grassmann variables $\phi_I$ with $I=1,2$, which are contracted.\footnote{As we will discuss later in section \ref{sec:forms}, it is natural to work in the $\mathcal{N}=4$ formalism for the construction of the orthogonal momentum amplituhedron, therefore $I=1,2$, instead of $I=1,2, 3$ in the case of the $\mathcal{N}=6$ formalism. This is realized through a SUSY reduction as we will show in detail in section \ref{sec:forms}.} On the space of $\Lambda$'s we define a kinematic bracket
\begin{align}
\langle  i_1 i_2 \ldots i_{k +2} \rangle = \epsilon_{A_1 A_2 \ldots A_{k+2} }  \Lambda^{A_1}_{i_1} \Lambda^{A_2}_{i_2} \ldots \Lambda^{A_{k +2}}_{i_{k +2} } \, .
\end{align}

We define the Orthogonal momentum amplituhedron as a Grassmannian element $Y$ given by:
\begin{align} \label{eq:Ymap}
Y^A_a = \sum_{i=1}^n  c_{a i} \Lambda^A_i ,
\end{align} 
where $a=1,\cdots,k$ and $A=1,\cdots,k+2$. Here $\Lambda^A_i $ is an element of the positive moment matrix, $c_{a i} $ is an element of positive orthogonal Grassmannian $OG_+(k,2k)$ in the positive branch.  The definition of the positive orthogonal Grassmannian which is the moduli space of null planes, as  discussed in \cite{Huang:2013owa, Huang:2014xza}. The important point is that the positive part of orthogonal Grassmannian is defined with respect to  the split signature metric $\eta^{ij}=(+,-,+,\dots,-)$, and the orthogonal constraints take the form:
\begin{align}
\label{eq:splitorthogonal}
\eta^{ij}C_{ai}C_{bj}=0\,.
\end{align}
In this signature, the minors satisfy $M_I/M_{\bar{I}}=\pm1$, where $\bar{I}$ is the ordered complement of $I$. For $M_I/M_{\bar{I}}=1(-1)$, the $OG_+(k,2k)$ is called ``positive (negative)" branch.

The dimension of the orthogonal momentum amplituhedron is $n-3$. First, since $Y\in G(k,k+2)$, we have:
\begin{align}
\text{dim}(G(k,k+2))= 2k=n\,.
\end{align} 
Indeed, the orthogonal momentum amplituhedron lives on a co-dimension 3 surface inside $G(k,k+2)$ satisfying:
\begin{equation}\label{eq:Orelation}
0=\sum_{i=1}^nP^{\alpha \beta}_i=\sum_{i=1}^n(-1)^i\left(Y^\perp \cdot \Lambda^T\right)^{\alpha}_i\left(Y^\perp \cdot \Lambda^T\right)^{\beta}_i.
\end{equation}
We can see this from the definition \eqref{eq:Ymap}. Let us start from the following equation:
\begin{equation}
0=Y^{\perp}\cdot Y^T=Y^{\perp}\cdot\Lambda^T\cdot C^T.
\end{equation}
Then the 2-dimensional space $Y^{\perp}\cdot\Lambda^T$ is a subspace of $(C^{T})^\perp$. This means that the space $Y^{\perp}\cdot\Lambda^T$ is orthogonal. Since we take the odd legs as the outgoing and even legs as ingoing, there is a $(-1)^i$ factor. Therefore the orthogonal momentum amplituhedron has dimension $n-3$.

Defining the planar Mandelstam variables to as
\begin{equation}
S_{i,i+1,\dots,i+p}=\sum_{i\leq j_1<j_2\leq i+p}(-1)^{j_1+j_2+1} \langle Y j_1j_2 \rangle^2\,,
\end{equation}
where the $(-1)$ factor reflects the fact that the odd (even) legs as outgoing (ingoing) momenta, the orthogonal momentum amplituhedron has two type of the boundaries:
\begin{equation}
\label{eq:oddboundary}
S_{\underbrace{i,i+1,\dots,i+p}_{\text{odd}}}=0,\qquad p=2, 4, 6, \dots \, ,
\end{equation}
\begin{equation}
\label{eq:evenboundary}
S_{\underbrace{i,i+1,\dots,i+p}_{\text{even}}}=0,\qquad p=1,3, 5, \dots \, .
\end{equation}
Note that since $\langle Y i\, i{+}1\rangle^2=S_{i,i+1}$, the boundary associated with $\langle Y i\, i{+}1\rangle=0$ is the same as $S_{i,i+1}$. As we will see later, only ``odd-particle Mandelstam variables" \eqref{eq:oddboundary} are the co-dimension one boundaries, the other ``even-particle Mandelstam variables" \eqref{eq:evenboundary}  are higher co-dimension boundaries. To see that these are boundaries, we need to first check the positivity of the planar Mandelstam variables for all points in the orthogonal momentum amplituhedron. Although this positivity is not manifest from the definition \eqref{eq:Ymap}, we can check this fact by using the explicit $C$-matrix parametrization. Here we have used the ``Veronese parametrization" of the $C$-matrx:\footnote{This is instrumental in connecting the geometry of the moduli space of punctured disk to the amplituhedron as explored recently in~\cite{Song}.}
\begin{equation}
C=\left(\begin{array}{ccccc}1 & 1 &1 & \cdots & 1 \\ t_1z_1 & t_2z_2 &t_3z_3 & \cdots & t_nz_n \\ t_1z_1^2 & t_2z_2^2 &t_3z_3^2 & \cdots & t_nz_n^2\\ \vdots & \vdots &\vdots & \cdots & \vdots \\t_1z_1^{k-1} & t_2z_2^{k-1} &t_3z_3^{k-1} & \cdots & t_nz_n^{k-1}\end{array}\right) \, ,
\end{equation}
where $t_2,\dots,t_n$ are given as
\begin{equation}
t_i=t_1\frac{\sqrt{\prod_{j\geq2,j\neq i}(z_j-z_1)}}{\sqrt{\prod_{j\geq2,j\neq i}(z_j-z_i)}}\, .
\end{equation}
When these parameters satisfy $t_1>0, z_i>0$ and $z_i-z_j>0 $ for $i>j$, all ordered minors of $C$ are positive. By using this parametrization, we have checked numerically that $S_{i,i+1,\dots,i+p}$ are indeed positive up to 10-points. 

In the section \ref{sec:forms}, we will further show that the BCFW cells tile the space of $Y$, and hence the boundaries in  (\ref{eq:oddboundary}) and  (\ref{eq:evenboundary}) are simply a reflection of that for the collection of cells.

\subsection{Sign flip definition}
The orthogonal momentum amplituhedron defined in  (\ref{eq:Ymap}) can also be carved out by imposing constraints directly on $Y$ through a set of sign flip conditions. First note that \eqref{eq:Ymap} is the same as the one of the ordinary amplituhedron with $m=2$ \cite{Arkani-Hamed:2013jha} except the orthogonal condition in $C$-matrix. Following \cite{Arkani-Hamed:2017vfh}, the sign flip definition of the $m=2$ amplituhedron
\begin{align}
\label{eq:signflip}
&\langle Y ii+1 \rangle >0 \, , \nonumber\\
&\{\langle Y12 \rangle, \langle Y13 \rangle,\dots,\langle Y1n \rangle\}\ \text{has $k$ sign flips}\, ,
\end{align} 
along with positive external data (in the sense of positive ordered minors) is conjectured to fix $Y$ to \eqref{eq:Ymap}. The additional condition is the orthogonality of the Grassmannian. We will show that the condition 
\begin{align}
\label{eq:signfliporthogonal}
\sum_{j=1}^n \,(-1)^j\,\langle Y j a  \rangle\langle Y j b  \rangle=0, \quad  \text{for}\; a, b=1,\cdots,n\, ,
\end{align} 
is equivalent to the orthogonal condition of the $C$-matrix. Therefore the sign flip conditions \eqref{eq:signflip} and the condition \eqref{eq:signfliporthogonal} give the sign flip defintion of the orthogonal momentum amplituhedron. This definition reveals the fact the geometry for $Y$ is the same as $\tilde{Y}$ for $\mathcal{N}=4$ sYM, with the extra orthogonal condition in \eqref{eq:signfliporthogonal}. Thus the geometry for ABJM amplitude lives on a subspace of the geometry for the split-helicity sector of $\mathcal{N}=4$ sYM!

To see the equivalence of \eqref{eq:signfliporthogonal} and the orthogonality of $C$-matrix,  we rewrite the relation \eqref{eq:signfliporthogonal} as
\begin{equation}
\sum_{j=1}^n(-1)^j\langle Y j a  \rangle\langle Y j b  \rangle=\sum_{j=1}^n\sum_{\substack{i_1<\dots<i_k\\j_1<\dots<j_k}}(-1)^jM_{i_1i_2\dots i_k}M_{j_1j_2\dots j_k}\langle i_1i_2\dots i_k ja \rangle \langle j_1j_2\dots j_k jb \rangle,
\end{equation}
where $M_{i_1,\dots, i_k}$ is the minor of the $C$-matrix. Let us consider only terms that are proportional to $\langle a_1a_2\dots a_k a_{k+1}a \rangle \langle b_1b_2\dots b_k b_{k+1}b \rangle$, where $ j\in (a_1,\dots,a_{k+1}),(b_1,\dots,b_{k+1})$.  These terms can be expressed as 
\begin{equation}\label{eq:Orelation3}
\sum_{j=c_1}^{c_{k'}}(-1)^jM_{a_1\dots\hat{j}\dots a_{k+1}}M_{b_1\dots\hat{j}\dots b_{k+1}}\langle a_1\dots \hat{j}\dots a_{k+1} ja \rangle \langle b_1\dots\hat{j}\dots b_{k+1}jb \rangle,
\end{equation}
where the sum of $j$ runs only the common parts $(a_1,\dots,a_{k+1})\cap(b_1,\dots,b_{k+1})\equiv (c_1,\dots,c_{k'})$ and $c_1=\max\{a_1,b_1\}, c_{k'}=\min\{a_k,b_k\} $. Without loss of generality, we can fix $\max\{a_1,b_1\}=a_1=b_m$ (for some integer $m$) and $\min\{a_k,b_k\}=a_k(=b_{k+m-1})$. Then the all kinematic brackets of the right side of \eqref{eq:Orelation3} become $(-1)^m\langle a_1\dots a_{k+1} a \rangle \langle b_1\dots b_{k+1} b \rangle$. Therefore equation \eqref{eq:Orelation3} reduces to
\begin{equation}
(-1)^m\left(\sum_{j=c_1}^{c_{k'}}(-1)^jM_{a_1\dots\hat{j}\dots a_{k+1}}M_{b_1\dots\hat{j}\dots b_{k+1}}\right)\langle a_1\dots a_{k+1} a \rangle \langle b_1\dots b_{k+1} b \rangle .
\end{equation}
Since the minors of the positive orthogonal Grassmannian satisfy $M_I=M_{\bar{I}}$, where $\bar{I}$ is the complement of $I$, therefore, 
\begin{equation}\label{eq:Orelation4}
\sum_{j=c_1}^{c_{k'}}(-1)^jM_{a_1\dots\hat{j}\dots a_{k+1}}M_{b_1\dots\hat{j}\dots b_{k+1}}=\sum_{j=c_1}^{c_{k'}}(-1)^jM_{a_1\dots\hat{j}\dots a_{k+1}}M_{\bar{b}_1\dots j\dots \bar{b}_{k-1}}=0.
\end{equation}
Here we used the Pl$\ddot{\text{u}}$cker relation
\begin{equation}
\sum_{l=1}^{k+1}(-1)^lM_{i_1\dots i_{k-1},j_l}M_{j_1\dots \hat{j_l}\dots j_{k+1}}=0.
\end{equation}
We conclude that, since our choice of the $(i_1,\dots,i_k,j),(j_1,\dots,j_k,j)$ is general, the relation \eqref{eq:signfliporthogonal} holds for general kinematics when the $C$-matrix satisfies the orthogonal conditions.

We further argue that (\ref{eq:Orelation4}) implies orthogonality in a similar manner as Appendix A of \cite{Huang:2013owa}. We gauge fix the $k \times 2k$ matrix to be 
\begin{equation}
C=(\mathbb{1},c)=\left(\begin{array}{ccccccc}1 & 0 & 0 & 0 & c_{1, k+1} & \cdots & c_{1,2k} \\0 & 1 & 0 & 0 & c_{2, k+1} & \cdots & c_{2,2k} \\0 & 0 & \cdots & 0 & \vdots & \vdots & \vdots \\0 & 0 & 0 & 1 & c_{k, k+1} & \cdots & c_{k,2k}\end{array}\right)\, .
\label{eq:GaugeApp}
\end{equation}
The orthogonal condition $C \cdot C^T=0 $ is equivalent to 
\begin{equation} \label{eq:c_dot_cT}
c \cdot c^T= c^T \cdot c= -\mathbb{1} \,. 
\end{equation}
While $c^T \cdot c= -\mathbb{1}$ gives
\begin{equation}\label{eq:cT_dot_c}
1+\sum_{j=1}^{k}c_{j, A}^2=0, \quad {\rm and} \; \sum_{j=1}^{k} c_{j, A}c_{j,B}=0 \,,  \quad A,B=k+1,\cdots, 2k\,,
\end{equation}
which we are going to show that it is equivalent to (\ref{eq:Orelation4}).
Let us first choose $\{a_1,\cdots,a_{k+1}\}=\{b_1,\cdots,b_{k+1}\}$. Without loss of generality, if we choose $\{a_1,\cdots,a_{k+1}\}$ to be the first $k+1$ columns of (\ref{eq:GaugeApp}), it is easy to see (\ref{eq:Orelation4}) gives the diagonal part of orthogonal constraints in (\ref{eq:cT_dot_c}). Next, we consider $\{a_1,\cdots,a_{k}\}=\{b_1,\cdots,b_{k}\}$ while  $a_{k+1}\neq b_{k+1}$. With no loss of generality we set
$\{a_1,\cdots,a_{k}\}$ to be the first $k$ columns of (\ref{eq:GaugeApp}), then one can see that in such choice, (\ref{eq:Orelation4}) produces the off-diagonal part of (\ref{eq:cT_dot_c}). 

This finishes the proof that the conditions \eqref{eq:signfliporthogonal} are equivalent to the orthogonal conditions on the $C$-matrix.

\section{Canonical forms from the Orthogonal Grassmannian}
\label{sec:forms}
In the previous section we have defined the orthogonal momentum amplituhedron as a positive kinematic map from the positive orthogonal Grassmannian. In this section we will consider its boundaries in more detail, showing that it corresponds to the physical boundaries of ABJM amplitude. Note that since the four and six-point amplitude corresponds to the top cell of $OG_{+}(2,4)$ and $OG_{+}(3,6)$ respectively, the boundary of the amplitude trivially matches to the amplituhedron. For more than six points, the amplitude is associated with a sum over lower dimensional cells (BCFW cells). Thus if the BCFW cells tile the amplituhedron, and are non-overlapping, then the boundaries of the amplitude can be mapped to those of the amplituhedron.  

Let us consider the first non-trivial example, the eight-point amplitude, which is a sum of two BCFW cells. We begin by choosing the top-cell $C$-matrix and fixed positive kinematics $\Lambda$. This gives a point $Y = C\cdot \Lambda$ inside the orthogonal momentum amplituhedron. We can check whether or not only one of the BCFW cells contains this point. More precisely, if we represent this point by using the BCFW $C$-matrices of the BCFW cells and the same positive kinematics, only one of them can reproduce this point. By checking this holds for many points inside the eight-point space numerically, we have verified that the eight-point BCFW cells are non-overlapping and tilling the orthogonal momentum amplituhedron space.

After confirming that the BCFW cells indeed tile the space, we can then utilize this connection to construct the volume form. We will construct the canonical form derived from the Grassmannian integral with reduced  SUSY.  We begin by discussing the $\mathcal{N}=4$ formalism of ABJM amplitudes and the corresponding orthogonal Grassmannian and on-shell diagram constructions.  We find the volume form of each on-shell diagram in the $\mathcal{N}=4$ formalism is naturally a canonical $d\log$ form. In contrast, for the case of $\mathcal{N}=6$ formalism, one needs to introduce the so-called Jacobian factors to incorporate the mismatch of the bosonic and fermionic delta-functions.  We will then study the canonical forms in the language of the orthogonal momentum amplituhedron.

\subsection{ABJM amplitudes and the Orthogonal Grassmannian in $\mathcal{N}=4$ formalism}

The ABJM theory is a three-dimensional Chern-Simons matter theory with $\mathcal{N}=6$ supersymmetry. The physical degrees of freedom consist of the 4 complex scalars $X_A$, 4 complex fermions $\psi^{A\alpha}$ and their complex conjugates $\bar{X}^A, \bar{\psi}_{A\alpha}$ with $A=1,2,3,4$ and  $\alpha=1,2$. These fields transform in the fundamental or anti-fundamental of the R-symmetry $SU(4)$ and in the bi-fundamental representation under the gauge group $U(N)\times U(N)$. The index $\alpha$ denotes the spinor representation in the three-dimensional Lorentz group.
Let us define super-fields of the ABJM 
\begin{align}
\Phi^{\mathcal{N}=6} =& X_4 + \eta_A \psi^A -\frac{1}{2}\epsilon^{ABC} \eta_A\eta_B X_C- \eta_1\eta_2\eta_3 \psi^4 \, ,\\
\bar{\Psi}^{\mathcal{N}=6}  =&  \bar{\psi}_4 + \eta_A \bar{X}^A -\frac{1}{2}\epsilon^{ABC} \eta_A\eta_B \bar{\psi}_C- \eta_1\eta_2\eta_3 \bar{X}^4,
\end{align}
here we have decomposed the fields as $X_A\rightarrow (X_4,X_A)$ and $\psi_A\rightarrow (\psi_4,\psi_A)$. 

The tree-level super-amplitudes in ABJM theory can be written as:
 \begin{align}
\mathcal{A}_n^{\text{tree}}=\delta^3(\sum_{i=1}^np_i)\delta^6(\sum_{i=1}^nq_i)F_n(\lambda_i,\eta_i)\, ,
\end{align}
where $p$ and $q$ are the on-shell momentum and supermomentum 
 \begin{align}
(p_i)^{\alpha \, \beta}=\lambda^{\alpha}_i\lambda^{\beta}_i,\ \ q^{\alpha\, A}_i=\lambda^{\alpha}_i\eta^A_i. 
\end{align}
The function $F_n$ is a rational function of Lorentz invariants. 

In three dimensions, the on-shell variables transform under the little group $Z_2$ as
\begin{align}\label{eq:littlegroup}
\lambda_i^\alpha \rightarrow -\lambda_i^\alpha,\ \ \ \eta_i^A \rightarrow -\eta_i^A. 
\end{align}
There are only two states, the fermion state that obtains a minus sign under \eqref{eq:littlegroup},  and the scalar state that does not. Under little group transformations \eqref{eq:littlegroup} if external leg $i$, the function $F_n$ changes as
\begin{equation}
F_n \rightarrow\begin{cases}
F_n\ \   i\in\Phi&\\
-F_n\ \ i\in\Psi .&
\end{cases}
\end{equation}
From this, there are only two classes of amplitudes
\begin{align}
\mathcal{A}_n(\bar{1}2\bar{3}\dots 2k),\ \ \mathcal{A}_n(1\bar{2}3\dots \bar{2k}),
\end{align}
here we denote $\bar{i}$ that leg $i$ is $\bar{\Psi}$ and use the fact that only even-multiplicity scattering amplitudes can be non-vanishing for this ABJM theory. 

As we remarked earlier, that it is vital to work in the $\mathcal{N}=4$ formalism for the construction of the orthogonal momentum amplituhedron.  One may obtain the $\mathcal{N}=4$ superfields from the more familiar $\mathcal{N}=6$ superfields through a SUSY reduction. They are defined as, 
\begin{align} \label{eq:N=4}
\Phi^{\mathcal{N}=4}  &:=  \Phi^{\mathcal{N}=6} \big{|}_{\eta^3 \rightarrow 0}  =X_4 + \eta_I \psi^I + (\eta)^2 X_3 \, , \cr
  \bar{\Phi}^{\mathcal{N}=4}  &:=\int d\eta^3 \bar{\Psi }^{\mathcal{N}=6} = \bar{X}^3 + \eta_I \bar{\psi}^I - (\eta)^2 \bar{X}^4 \, , \cr
  {\Psi}^{\mathcal{N}=4}  &:=\int d\eta^3   \Phi^{\mathcal{N}=6}  = {\psi}^3 + \eta_I  {X}^I + (\eta)^2 {\psi}^4 \, , \cr
\bar{\Psi}^{\mathcal{N}=4}  &:= \bar{\Psi }^{\mathcal{N}=6}   \big{|}_{\eta^3 \rightarrow 0}  = \bar{\psi}_4 + \eta_I \bar{X}^I + (\eta)^2 \bar{\Psi}_3  \, ,
 \end{align}
 here we have decomposed the fields as $X_A\rightarrow (X_4,X_3,X_I)$ and $\psi_A\rightarrow (\psi_4,\psi_3,\psi_I)$, with $I=1,2$.  The superamplitudes in the $\mathcal{N}=4$ formalism can again be obtained by the same SUSY reduction, namely setting $k$ of $\eta^3 \rightarrow 0$ and integrating out the other $k$ of $\eta^3$ for a $2k$-point superamplitude.

The orthogonal Grassmannian is defined as the space of $k$-planes in $\mathbb{C}^n$, such that $\eta^{ij}C_{ai}C_{bj}=0$. A tree-level $(n=2k)$-point scattering amplitudes of ABJM is given as a sum of the residues of the integral over an orthogonal Grassmannian $C_{ai}\in OG(k,2k)$
\begin{align} \label{eq:N4OG}
\mathcal{A}_{2k} = \int\frac{d^{2k\times k} C_{ai}}{\text{Vol}(\text{GL}(k))}\frac{M_{i_1, i_2,\dots, i_k}}{M_1 M_{2}\dots M_{k}}\delta^{k(k+1)/2}(C\cdot C^T)\prod_{a=1}^k \delta^{2|2}(C_a\cdot \Lambda),
\end{align}
where $M_i$ are the $i$-th consecutive minor
\begin{align} 
M_i\equiv \sum_{a_1, a_2, \dots, a_k}\epsilon_{a_1 a_2 \dots a_k} c_{ia_1} c_{i+1a_2}\dots c_{i+k a_k}  \, . 
\end{align}
The numerator $M_{i_1, i_2,\dots, i_k}$ is given by
\begin{align} 
M_{i_1, i_2,\dots, i_k} =  \sum_{a_1, a_2, \dots, a_k}\epsilon_{a_1 a_2 \dots a_k} c_{i_1 a_1} c_{i_2 a_2}\dots c_{i_k a_k} \, . 
\end{align}
It is due to the fact that we work in the  $\mathcal{N}=4$ formalism, arising from the SUSY reduction we discussed above, where $i_1, i_2, \dots, i_k$ are superfields of either $\bar{\Phi}^{\mathcal{N}=4}$  or  ${\Psi}^{\mathcal{N}=4}$ in \eqref{eq:N=4}. The integration over $\eta^3$ for each of these fields generates $M_{i_1, i_2,\dots, i_k}$. 

A few remarks are in order here.  Firstly, in the $\mathcal{N}=4$ formalism, as indicated in \eqref{eq:N4OG}, the bosonic and fermionic delta-functions, $\delta^{2|2}(C_a\cdot \Lambda)$, match each other. One of the consequences of this is that, unlike $\mathcal{N}=6$ formalism \cite{Huang:2013owa, Huang:2014xza}, the so-called Jacobi are not required for the volume forms of the on-shell diagrams in the $\mathcal{N}=4$ formalism.  As we will see shortly, they are given by products of canonical $d\log$ forms for each on-shell diagram.  Secondly, due to the fact that the geometry of orthogonal momentum amplituhedron has the cyclic invariance, we will consider the amplitudes 
\begin{align} \label{eq:N4amp}
\mathcal{A}_{2k}( \bar{\Phi}^{\mathcal{N}=4}_1, \Phi^{\mathcal{N}=4}_2, \bar{\Phi}^{\mathcal{N}=4}_3, \Phi^{\mathcal{N}=4}_4, \dots, \Phi^{\mathcal{N}=4}_{2k}) \, ,
\end{align}
which implies the numerator in \eqref{eq:N4OG} is $M_{1, 3,\dots, 2k{-}1}$\footnote{One may also consider $\mathcal{A}_{2k}( \bar{\Psi}^{\mathcal{N}=4}_1, \Psi^{\mathcal{N}=4}_2, \bar{\Psi}^{\mathcal{N}=4}_3, \Psi^{\mathcal{N}=4}_4, \dots, \Psi^{\mathcal{N}=4}_{2k})$, for which we have $M_{2, 4,\dots, 2k}$ in the numerator.}.  With this choice, $\mathcal{A}_{2k}$ defined in \eqref{eq:N4OG} has the cyclic invariance.  Thirdly, again thanks to the match of bosonic and fermionic variables in the   $\mathcal{N}=4$ formalism, one may identify $\eta^I$ by $d\lambda^{\alpha}$. This will lead to a differential form representation of scattering amplitudes in ABJM theory in an analogous construction of scattering amplitudes in ${\mathcal{N}=4}$ sYM in four dimensions \cite{He:2018okq}.

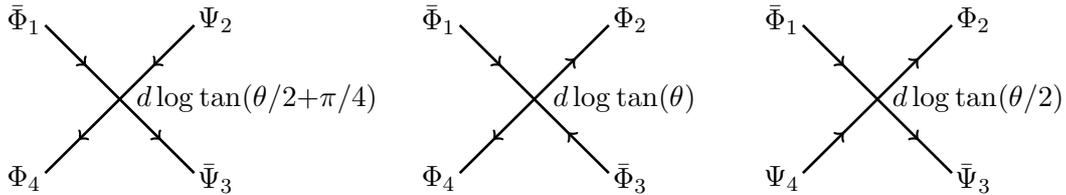
\begin{figure}
\begin{center} 
		\begin{tikzpicture}[scale=0.7, line width=1 pt]
	\draw [3Dline] (1.4,1.4)--(0,0);
	\draw [3Dline] (0,0)--(1.4,-1.4);
	\draw [3Dline] (-1.4,1.4)--(0,0);
	\draw [3Dline] (0,0)--(-1.4,-1.4);
	\node at (2.4,0) {$~~d\log \tan(\theta/2 {+}\pi/4)$};
	\node at (-1.8, 1.5) {$\bar{\Phi}_1$};
	\node at (1.8, 1.5) {${\Psi}_2$};
		\node at (1.8, -1.5) {$\bar{\Psi}_3$};
			\node at (-1.8, -1.5) {$\Phi_4$};
			\end{tikzpicture} 
	$\,$
	\begin{tikzpicture}[scale=0.7, line width=1 pt]
	\draw [3Dline] (0,0)--(1.4,1.4);
	\draw [3Dline] (1.4,-1.4)--(0,0);
	\draw [3Dline] (-1.4,1.4)--(0,0);
	\draw [3Dline] (0,0)--(-1.4,-1.4);
	\node at (1.6,0) {$~d\log \tan(\theta)$};
	\node at (-1.8, 1.5) {$\bar{\Phi}_1$};
	\node at (1.8, 1.5) {${\Phi}_2$};
		\node at (1.8, -1.5) {$\bar{\Phi}_3$};
			\node at (-1.8, -1.5) {$\Phi_4$};
	\end{tikzpicture}
	$\quad $
\begin{tikzpicture}[scale=0.7, line width=1 pt]
	\draw [3Dline] (0,0)--(1.4,1.4);
	\draw [3Dline] (0,0)--(1.4,-1.4);
	\draw [3Dline] (-1.4,1.4)--(0,0);
	\draw [3Dline] (-1.4,-1.4)--(0,0);
	\node at (1.8,0) {$~d\log \tan(\theta/2)$}; 
	\node at (-1.8, 1.5) {$\bar{\Phi}_1$};
	\node at (1.8, 1.5) {${\Phi}_2$};
		\node at (1.8, -1.5) {$\bar{\Psi}_3$};
			\node at (-1.8, -1.5) {$\Psi_4$};
		\end{tikzpicture}  
\end{center}
\caption{The three four-point vertices of $\mathcal{N}=4$ formalism of on-shell diagram representation of scattering amplitudes in ABJM theory.}
\label{fig:4pt_diagrams}
\end{figure}

The building blocks of on-shell diagrams for ABJM theory are the four-point amplitudes, which in the $\mathcal{N}=4$ formalism are given by
\begin{align} \label{eq:4-point}
\mathcal{A}_4 ( \bar{\Phi}^{\mathcal{N}=4}_1, \Psi^{\mathcal{N}=4}_2, \bar{\Psi}^{\mathcal{N}=4}_3, \Phi^{\mathcal{N}=4}_4) &= \int dC {\delta^3 (C C^T) \delta^{2|2}(C \cdot \Lambda)  } {1 \over (23)}  \, ,  \cr
\mathcal{A}_4 ( \bar{\Phi}^{\mathcal{N}=4}_1, \Phi^{\mathcal{N}=4}_2, \bar{\Phi}^{\mathcal{N}=4}_3, \Phi^{\mathcal{N}=4}_4) &= \int dC {\delta^3 (C C^T) \delta^{2|2}(C \cdot \Lambda)  } {(13) \over (12)(23) }  \, , \cr
\mathcal{A}_4 ( \bar{\Phi}^{\mathcal{N}=4}_1, \Phi^{\mathcal{N}=4}_2, \bar{\Psi}^{\mathcal{N}=4}_3, \Psi^{\mathcal{N}=4}_4) & =\int dC {\delta^3 (C C^T) \delta^{2|2}(C \cdot \Lambda)  } {(14) \over (12)(23)} \, .
\end{align}
Using the $OG_+(2,4)$, 
\begin{align}
C= \begin{pmatrix}
1&\cos\theta&0&-\sin\theta\\
0&\sin\theta&1&\cos\theta
\end{pmatrix} \, ,
\end{align} 
we find the three types of four-point amplitudes can all be expressed in $d\log$ forms, as shown in Fig.\ref{fig:4pt_diagrams}. The incoming arrows represent the superfields $\bar{\Phi}^{\mathcal{N}=4}$ or $\Psi^{\mathcal{N}=4}$, and they are obtained by integrating out $\eta^3$ as shown in \eqref{eq:N=4}; whereas the outgoing arrows represent the superfields ${\Phi}^{\mathcal{N}=4}$ or $\bar{\Psi}^{\mathcal{N}=4}$, which are obtained by setting $\eta^3 \rightarrow 0$. 

These four-point vertices form building blocks for the on-shell diagrams of the amplitudes in ABJM theory, and one may glue them together to form more general diagrams.   Generally, the $n$-point superamplitudes in the $\mathcal{N}=4$ formalism can be expressed as
\begin{align} \label{eq:n-point}
\mathcal{A}_{2k}  = \int { \prod_{a=1}^{k} \delta^{2|2}(C_a \cdot \Lambda)  } \, \omega_{2k} \, ,
\end{align}
here $n=2k$. The integrand $\omega_{2k}$ is obtained by gluing the four-point vertices given in Fig.\ref{fig:4pt_diagrams} in all possible ways following the BCFW construction of tree-level amplitudes. Each diagram is given by products of $d\log$'s, and $\omega_{2k}$ is a sum of these canonical $d\log$ forms.  As we anticipated, when we express $c_{ai}$ of $OG_{+}(k, 2k)$ in terms of $Y$ under the support of $Y = C \cdot \Lambda$, $\omega_{2k}$ essentially becomes the canonical form of the orthogonal momentum amplituhedron, $\bold {\Omega}^{3d}_{2k,k}$, which we will study in details in the next section. It is therefore vital to construct $\omega_{2k}$, as we will do below.

\begin{figure}
\begin{center} 
	\begin{tikzpicture}[scale=1, line width=1 pt]
	\draw [3Dline] (0,0.4)--(-1.2,-1.4);
		\draw [3Dline] (-1.2,-1.4)--(1.2,-1.4);
		\draw [3Dline] (1.2,-1.4)--(0,0.4);
		\draw [3Dline] (0,0.4)--(0.5,1);
		\draw [3Dline] (-0.5,1)--(0,0.4);
		\draw [3Dline] (1.9,-1.4)--(1.2,-1.4);
		\draw [3Dline] (1.2,-1.4)--(1.65,-2.1);
		\draw [3Dline] (-1.65,-2.1)--(-1.2,-1.4);
		\draw [3Dline] (-1.2,-1.4)--(-1.9,-1.4);
		\node at (0.4,0.4) {$\theta_1$};
		\node at (1.4,-1) {$\theta_2$};
		\node at (-1.4,-1) {$\theta_3$};
		\node at (-0.7,1.2) {$1$};
		\node at (-2.3,-1.4) {$6$};
		\node at (0,-2.4) {(a)};
	\end{tikzpicture}
	$\qquad \qquad$
	\begin{tikzpicture}[scale=1, line width=1 pt]
	\draw [3Dline] (-1.2,-1.4)--(0,0.4);
		\draw [3Dline] (1.2,-1.4)--(-1.2,-1.4);
		\draw [3Dline] (0,0.4)--(1.2,-1.4);
		\draw [3Dline] (0,0.4)--(0.5,1);
		\draw [3Dline] (-0.5,1)--(0,0.4);
		\draw [3Dline] (1.9,-1.4)--(1.2,-1.4);
		\draw [3Dline] (1.2,-1.4)--(1.65,-2.1);
		\draw [3Dline] (-1.65,-2.1)--(-1.2,-1.4);
		\draw [3Dline] (-1.2,-1.4)--(-1.9,-1.4);
		\node at (0.4,0.4) {$\theta_1$};
		\node at (1.4,-1) {$\theta_2$};
		\node at (-1.4,-1) {$\theta_3$};
		\node at (-0.7,1.2) {$1$};
		\node at (-2.3,-1.4) {$6$};
		\node at (0,-2.4) {(b)};
		\end{tikzpicture}  
\end{center}
\caption{The two six-point diagrams correspond to different ways of arranging arrow flow of the internal lines. Here and throughout this paper we only label the first and last external legs.}
\label{fig:6pt_diagrams}
\end{figure}
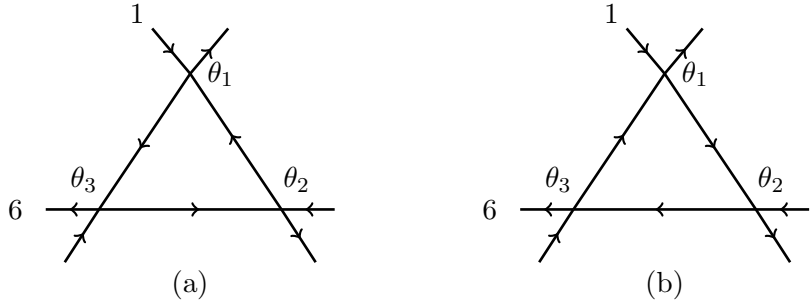

Let us begin with the six-point case as an example. There are two diagrams contributing to six-point amplitude, as shown in Fig.\ref{fig:6pt_diagrams} that correspond to two different choices of internal arrow flows. Each diagram in this formalism takes a canonical $d\log$ form. The contribution from the diagram on the left, Fig.\ref{fig:6pt_diagrams} (a), is given by
    \begin{equation}
\omega_{6,1} = \bigwedge_{i=1}^3 d \log \tan  (\theta_i)\, ,
    \end{equation}
    and the contribution from the right, Fig.\ref{fig:6pt_diagrams} (b), can be expressed as
    \begin{equation}
\omega_{6,2} = \bigwedge_{i=1}^3 d \log \tan (\theta_i/2) =  c_1 c_2 c_3  \bigwedge_{i=1}^3 d \log \tan  (\theta_i) \,. 
    \end{equation}
They are obtained by simply gluing the four-point amplitudes given in \eqref{eq:4-point} and  Fig.  \ref{fig:4pt_diagrams}.  One may combine these two contributions, which lead to
    \begin{equation} \label{eq:6pt_dlog}
\omega_{6} =\omega_{6,1}+\omega_{6,2}  =     (1+c_1 c_2 c_3) \bigwedge_{i=1}^3 d \log \tan \theta_i\,.  
    \end{equation}
    This is in agreement with the result in \cite{Huang:2013owa, Huang:2014xza} using the $\mathcal{N}=6$ formalism.  In the $\mathcal{N}=6$ formalism, there is a single BCFW diagram, due to the mismatch of the bosonic and fermionic delta-functions, which leads to  the prefactor $(1+c_1 c_2 c_3)$ arising as a Jacobian. 

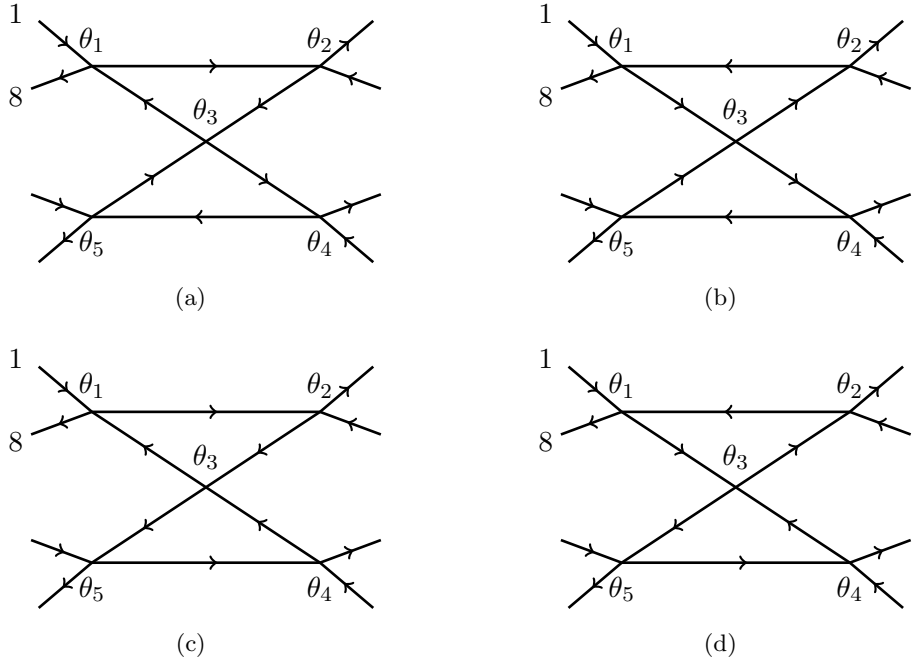
\begin{figure}
    \centering
    \subfigure[]{\begin{tikzpicture}[scale=1, line width=1 pt]
	\draw [3Dline] (1.5,1)--(0,0);
	\draw [3Dline] (0,0)--(-1.5,1);
	\draw [3Dline] (-1.5,1)--(1.5,1);
	\draw [3Dline] (0,0)--(1.5,-1);
	\draw [3Dline] (-1.5,-1)--(0,0);
	\draw [3Dline] (1.5,-1)--(-1.5,-1);
    \draw [3Dline] (-2.2,1.6)--(-1.5,1);
	\draw [3Dline] (-1.5,1)--(-2.3,0.7);
	\draw [3Dline] (1.5,1)--(2.2,1.6);
	\draw [3Dline] (2.3,0.7)--(1.5,1);
	\draw [3Dline] (-1.5,-1)--(-2.2,-1.6);
	\draw [3Dline] (-2.3,-0.7)--(-1.5,-1);
	\draw [3Dline] (2.2,-1.6)--(1.5,-1);
	\draw [3Dline] (1.5,-1)--(2.3,-0.7);
		\node at (-1.5,1.35) {$\theta_1$};
		\node at (1.5,1.35) {$\theta_2$};
		\node at (0,0.4) {$\theta_3$};
		\node at (-1.5,-1.35) {$\theta_5$};
		\node at (1.5,-1.35) {$\theta_4$};
		\node at (-2.5,1.7) {$1$};
		\node at (-2.5,0.6) {$8$};
		\end{tikzpicture} }  $\qquad \qquad$
    \subfigure[]{\begin{tikzpicture}[scale=1, line width=1 pt]
	\draw [3Dline] (0,0)--(1.5,1);
	\draw [3Dline] (-1.5,1)--(0,0);
	\draw [3Dline] (1.5,1)--(-1.5,1);
	\draw [3Dline] (0,0)--(1.5,-1);
	\draw [3Dline] (-1.5,-1)--(0,0);
	\draw [3Dline] (1.5,-1)--(-1.5,-1);
    \draw [3Dline] (-2.2,1.6)--(-1.5,1);
	\draw [3Dline] (-1.5,1)--(-2.3,0.7);
	\draw [3Dline] (1.5,1)--(2.2,1.6);
	\draw [3Dline] (2.3,0.7)--(1.5,1);
	\draw [3Dline] (-1.5,-1)--(-2.2,-1.6);
	\draw [3Dline] (-2.3,-0.7)--(-1.5,-1);
	\draw [3Dline] (2.2,-1.6)--(1.5,-1);
	\draw [3Dline] (1.5,-1)--(2.3,-0.7);
			\node at (-1.5,1.35) {$\theta_1$};
		\node at (1.5,1.35) {$\theta_2$};
		\node at (0,0.4) {$\theta_3$};
		\node at (-1.5,-1.35) {$\theta_5$};
		\node at (1.5,-1.35) {$\theta_4$};
		\node at (-2.5,1.7) {$1$};
		\node at (-2.5,0.6) {$8$};
		\end{tikzpicture} } \\
    \subfigure[]{\begin{tikzpicture}[scale=1, line width=1 pt]
	\draw [3Dline] (1.5,1)--(0,0);
	\draw [3Dline] (0,0)--(-1.5,1);
	\draw [3Dline] (-1.5,1)--(1.5,1);
	\draw [3Dline] (1.5,-1)--(0,0);
	\draw [3Dline] (0,0)--(-1.5,-1);
	\draw [3Dline] (-1.5,-1)--(1.5,-1);
	\draw [3Dline] (-2.2,1.6)--(-1.5,1);
	\draw [3Dline] (-1.5,1)--(-2.3,0.7);
	\draw [3Dline] (1.5,1)--(2.2,1.6);
	\draw [3Dline] (2.3,0.7)--(1.5,1);
	\draw [3Dline] (-1.5,-1)--(-2.2,-1.6);
	\draw [3Dline] (-2.3,-0.7)--(-1.5,-1);
	\draw [3Dline] (2.2,-1.6)--(1.5,-1);
	\draw [3Dline] (1.5,-1)--(2.3,-0.7);
	
	\node at (-1.5,1.35) {$\theta_1$};
		\node at (1.5,1.35) {$\theta_2$};
		\node at (0,0.4) {$\theta_3$};
		\node at (-1.5,-1.35) {$\theta_5$};
		\node at (1.5,-1.35) {$\theta_4$};
		\node at (-2.5,1.7) {$1$};
		\node at (-2.5,0.6) {$8$};
		\end{tikzpicture} } $\qquad \qquad$
    \subfigure[]{\begin{tikzpicture}[scale=1, line width=1 pt]
	\draw [3Dline] (0,0)--(1.5,1);
	\draw [3Dline] (-1.5,1)--(0,0);
	\draw [3Dline] (1.5,1)--(-1.5,1);
	\draw [3Dline] (1.5,-1)--(0,0);
	\draw [3Dline] (0,0)--(-1.5,-1);
	\draw [3Dline] (-1.5,-1)--(1.5,-1);
	\draw [3Dline] (-2.2,1.6)--(-1.5,1);
	\draw [3Dline] (-1.5,1)--(-2.3,0.7);
	\draw [3Dline] (1.5,1)--(2.2,1.6);
	\draw [3Dline] (2.3,0.7)--(1.5,1);
	\draw [3Dline] (-1.5,-1)--(-2.2,-1.6);
	\draw [3Dline] (-2.3,-0.7)--(-1.5,-1);
	\draw [3Dline] (2.2,-1.6)--(1.5,-1);
	\draw [3Dline] (1.5,-1)--(2.3,-0.7);
		\node at (-1.5,1.35) {$\theta_1$};
		\node at (1.5,1.35) {$\theta_2$};
		\node at (0,0.4) {$\theta_3$};
		\node at (-1.5,-1.35) {$\theta_5$};
		\node at (1.5,-1.35) {$\theta_4$};
		\node at (-2.5,1.7) {$1$};
		\node at (-2.5,0.6) {$8$};
		\end{tikzpicture} }
    \caption{There are eight BCFW diagrams that contribute to the eight-point tree-level amplitudes. Here we have only listed four diagrams corresponding to different ways of arranging arrow flow of the internal lines, the other set of four diagrams can be obtained by a simple cyclic shift on the external particles. }
    \label{fig:8pt_diagrams}
\end{figure}

The construction applies to on-shell diagram representation of higher-point amplitudes, both at tree and loop levels.  We conclude this section by considering the eight-point BCFW diagrams, as shown in Fig.\ref{fig:8pt_diagrams}. Here we only show explicitly one set of four BCFW diagrams, there are four more diagrams, which can be obtained from those in Fig.\ref{fig:8pt_diagrams} by a cyclic shift. The contribution from each diagram in  Fig.\ref{fig:8pt_diagrams} again is given by a canonical $d\log$ form, 
\begin{align}
\omega_{8,1}&=\bigwedge_{i=1}^5 d \log \tan  (\theta_i) \,,         \\
  \omega_{8,2} &=\bigwedge_{i=1}^3 d \log \tan  (\theta_i/2{+}\pi/4) \bigwedge_{i=4}^5 d \log \tan  (\theta_i) =s_1 s_2 s_3\bigwedge_{i=1}^5 d \log \tan  (\theta_i) \,,              \cr
\omega_{8,3} &=  \bigwedge_{i=1}^2 d \log \tan  (\theta_i)  \bigwedge_{i=3}^5 d \log \tan  (\theta_i/2{+}\pi/4) =s_3 s_4 s_5\bigwedge_{i=1}^5 d \log \tan  (\theta_i) \,,        \cr
  \omega_{8,4}&=\bigwedge_{i=1}^2 d \log \tan  (\theta_i/2{+}\pi/4) \wedge d \log \tan  (\theta_3)  \bigwedge_{i=4}^5 d \log \tan  (\theta_i/2{+}\pi/4)=s_1 s_2 s_4 s_5\bigwedge_{i=1}^5 d \log \tan  (\theta_i)  \,.  \nonumber
\end{align}
Combining the above four contributions, we have 
\begin{equation}
\omega_{8} = (1+s_1 s_2 s_3+s_3 s_4 s_5+s_1 s_2 s_4 s_5) \bigwedge_{i=1}^5 d \log \tan \theta_i\,.  
    \end{equation}
The final expression agrees with the volume form of one of the BCFW diagrams  (there are two BCFW diagrams in the $\mathcal{N}=6$ formalism) for the eight-point amplitude obtained originally in \cite{Huang:2014xza} using $\mathcal{N}=6$ formalism. In particular, the prefactor $ (1+s_1 s_2 s_3+s_3 s_4 s_5+s_1 s_2 s_4 s_5)$ arises as a Jacobian due to the mismatch of the bosonic and fermionic delta-functions.

\subsection{The canonical forms and boundaries}
\label{sec:momentumamp}

In this section, we will construct the canonical forms in the $Y$ space of the amplituhedron for tree-level amplitudes in ABJM theory. 
The dimension of the orthogonal momentum amplituhedron is $(n-3)$, and the canonical form $\bold{\Omega}^{3d}_{2k,k}$ is also $(n-3)$ dimensional. We define the independent expression of the volume function $\Omega^{3d}_{2k,k}$ by using $1=\delta^3(P)d^3P$ as follows:
\begin{align} \label{eq:3d_volume_form}
\bold{\Omega}^{3d}_{2k,k}\wedge d^3P \delta^3(P)=\prod_{a=1}^{k}\langle Y_1\dots Y_{k}d^2Y_a \rangle\delta^3(P)\Omega^{3d}_{2k,k}\, .
\end{align} 

In the following, we demonstrate how to obtain volume function $\Omega^{3d}_{2k,k}$ from the canonical form $\bold{\Omega}^{3d}_{2k,k}$ through the definition (\ref{eq:3d_volume_form}) for $k=2,3$ (i.e. four- and six-point amplitudes).
For $k=2$, the canonical form associated with the amplitude \eqref{eq:N4amp} is given in \eqref{eq:4-point} (or Fig.\ref{fig:4pt_diagrams}), which is simply $d \log \tan  (\theta)$. Using $Y^A_a = \sum_{i=1}^{4} c_{ai} \Lambda_i^A$, we can recast the result in the $Y$ space, which leads to
\begin{align} 
\bold{\Omega}^{3d}_{4,2} =d \log \frac{\la Y12 \ra}{\la Y23 \ra} \, .
\end{align} 
Using the definition (\ref{eq:3d_volume_form}), we find that the volume function is given by
\begin{align} \label{eq:4pts}
\Omega^{3d}_{4,2}(Y, \Lambda)=  \frac{\la Y13 \ra }{\la Y12 \ra \la Y23 \ra}\, \la 1234 \ra^2 \,.
\end{align}
For the six-point case, the canonical form is given by (\ref{eq:6pt_dlog}). The relation between $\tan  (\theta_i)$ and Y-bracket can be explicitly solved according to the parametrization of $OG_+(3,6)$, recasting in the Y space we have
\begin{align}  \label{eq:6pt_bold_form}
\bold{\Omega}^{3d}_{6,3} = \,&\frac{1}{8} \times d \log \left[\left(\frac{A^+_{54}}{A^+_{36}}\right)^2-1 \right] \wedge d \log \left[\left(\frac{A^+_{16}}{A^+_{52}}\right)^2-1 \right] \wedge d \log \left[\left(\frac{A^+_{32}}{A^+_{14}}\right)^2-1 \right] \nonumber \\
+&\frac{1}{8} \times d \log \left[ \frac{A^+_{54}+A^+_{36}}{A^+_{54}-A^+_{36}} \right] \wedge  d \log \left[ \frac{A^+_{16}+A^+_{52}}{A^+_{16}-A^+_{52}} \right] \wedge  d \log \left[ \frac{A^+_{32}+A^+_{14}}{A^+_{32}-A^+_{14}} \right] \,.
\end{align} 
where $A_{ab}^{\pm}$ are defined as
\begin{align}
A_{ab}^{\pm}=\sum_{i=1,3,5}\la Yia\ra\la Yib \ra \pm \la Y a{+}2\, a{-}2\ra\la Yb{-}2\, b{+}2 \ra \, .
\end{align}
They are related to the three-particle planar Mandelstam variables as follows
\begin{align} \label{eq:6pt_half_pole}
A_{52}^{+}\,A_{52}^{-}=-S_{1,2,3}\, S_{1,3,5}\,,\;\;
A_{36}^{+}\,A_{36}^{-}=-S_{2,3,4}\, S_{1,3,5}\,,\;\;
A_{14}^{+}\,A_{14}^{-}=-S_{3,4,5}\, S_{1,3,5}\,.
\end{align}
Plugging the above canonical form (\ref{eq:6pt_bold_form}) in (\ref{eq:3d_volume_form}), we obtain the six-point volume function 
\begin{align} \label{eq:6pts}
\Omega^{3d}_{6,3}(Y, \Lambda)
= \frac{(\sum_{i,j=1,3,5}\la Y ij \ra\la ij246 \ra+(1,3,5)\leftrightarrow (2,4,6))^2 S_{1,3,5}}{A_{52}^+\, A_{36}^+ \, A_{14}^+}\,,
\end{align} 

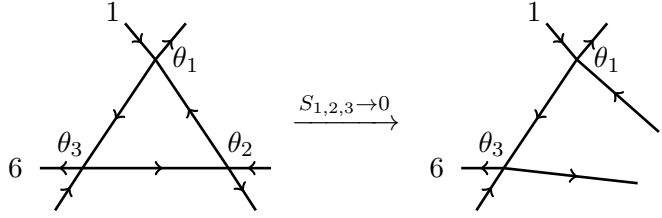
\begin{figure}
\begin{center} 
	\begin{tikzpicture}[scale=0.8, line width=1 pt]
	\draw [3Dline] (0,0.4)--(-1.2,-1.4);
		\draw [3Dline] (-1.2,-1.4)--(1.2,-1.4);
		\draw [3Dline] (1.2,-1.4)--(0,0.4);
		\draw [3Dline] (0,0.4)--(0.5,1);
		\draw [3Dline] (-0.5,1)--(0,0.4);
		\draw [3Dline] (1.9,-1.4)--(1.2,-1.4);
		\draw [3Dline] (1.2,-1.4)--(1.65,-2.1);
		\draw [3Dline] (-1.65,-2.1)--(-1.2,-1.4);
		\draw [3Dline] (-1.2,-1.4)--(-1.9,-1.4);
		\node at (0.5,0.4) {$\theta_1$};
		\node at (1.4,-1) {$\theta_2$};
		\node at (-1.4,-1) {$\theta_3$};
		\node at (-0.7,1.2) {$1$};
		\node at (-2.3,-1.4) {$6$};
		\end{tikzpicture}
		\begin{tikzpicture}[scale=0.8, line width=1 pt]
		\node at (0,1.5) {$\xrightarrow[]{S_{1,2,3}\rightarrow 0}$};
		\node at (0,0) {};
	\end{tikzpicture}
	\begin{tikzpicture}[scale=0.8, line width=1 pt]
	\draw [3Dline] (0,0.4)--(-1.2,-1.4);
		\draw [3Dline] (-1.2,-1.4)--(1.0,-1.65);
		\draw [3Dline] (1.35,-0.8)--(0,0.4);
		\draw [3Dline] (0,0.4)--(0.5,1);
		\draw [3Dline] (-0.5,1)--(0,0.4);
		\draw [3Dline] (-1.65,-2.1)--(-1.2,-1.4);
		\draw [3Dline] (-1.2,-1.4)--(-1.9,-1.4);
		\node at (0.5,0.4) {$\theta_1$};
		\node at (-1.4,-1) {$\theta_3$};
		\node at (-0.7,1.2) {$1$};
		\node at (-2.3,-1.4) {$6$};
		\end{tikzpicture}  
\end{center}
\caption{The diagram shows the factorization limit of a six-point diagram, $S_{1,2,3}\rightarrow 0$, which corresponds to taking $\theta_2 \rightarrow \pi/2$ on the diagram (a) in Fig.\ref{fig:6pt_diagrams}, and note that diagram (b) doesn't develop a singularity in this limit.}
\label{fig:6pt_fac}
\end{figure}

\begin{figure}
\begin{center} 
	\begin{tikzpicture}[scale=0.8, line width=1 pt]
	\draw [3Dline] (0,0.4)--(-1.2,-1.4);
		\draw [3Dline] (-1.2,-1.4)--(1.2,-1.4);
		\draw [3Dline] (1.2,-1.4)--(0,0.4);
		\draw [3Dline] (0,0.4)--(0.5,1);
		\draw [3Dline] (-0.5,1)--(0,0.4);
		\draw [3Dline] (1.9,-1.4)--(1.2,-1.4);
		\draw [3Dline] (1.2,-1.4)--(1.65,-2.1);
		\draw [3Dline] (-1.65,-2.1)--(-1.2,-1.4);
		\draw [3Dline] (-1.2,-1.4)--(-1.9,-1.4);
		\node at (0.5,0.4) {$\theta_1$};
		\node at (1.4,-1) {$\theta_2$};
		\node at (-1.4,-1) {$\theta_3$};
		\node at (-0.7,1.2) {$1$};
		\node at (-2.3,-1.4) {$6$};
		\end{tikzpicture}
		\begin{tikzpicture}[scale=0.8, line width=1 pt]
		\node at (0,1.5) {$\;\;\xrightarrow[]{\langle Y12 \rangle \rightarrow 0}$};
		\node at (0,0) {};
	\end{tikzpicture}
	\begin{tikzpicture}[scale=0.8, line width=1 pt]
		\draw [3Dline] (1.9,-1.4)--(1.2,-1.4);
		\draw [3Dline] (1.2,-1.4)--(1.65,-2.1);
		\draw [3Dline] (-1.65,-2.1)--(-1.2,-1.4);
		\draw [3Dline] (-1.2,-1.4)--(-1.9,-1.4);
		\node at (1.4,-1) {$\theta_2$};
		\node at (-1.4,-1) {$\theta_3$};
		\node at (-1.1,0.5) {$1$};
		\node at (-2.3,-1.4) {$6$};
				\path [draw=black,postaction={on each segment={mid arrow=black}}]
		(-0.85,0.4) to [bend right] (0.85,0.4)
  (-1.2,-1.4) to [bend right] (1.2,-1.4)
    (1.2,-1.4) to [bend right] (-1.2,-1.4);
		\end{tikzpicture}  
\end{center}
\caption{The diagram shows the soft limit of a six-point diagram, $\la Y12\ra \rightarrow0$, which corresponds to taking $\theta_1 \rightarrow \pi/2$ on the diagrams (a) or (b) in Fig.\ref{fig:6pt_diagrams}, here we show the case for diagram (a).}
\label{fig:6pt_soft}
\end{figure}
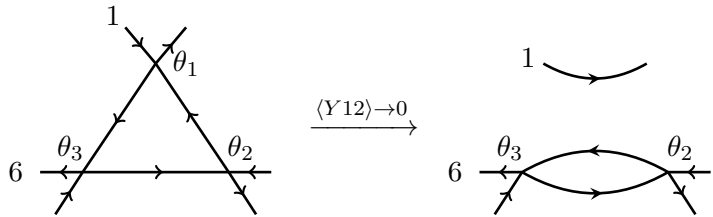

We conclude the discussions by studying the boundaries of the momentum amplituhedron. As we remarked previously that the planar Mandelstam variables are all positive for the positive Grassmannian and positive moment kinematics. The volume function at six points develops a singularity when $A_{52}$ approaches to zero, according to (\ref{eq:6pt_half_pole}) and note that $A_{ab}^-$ never vanish in the positive region, $S_{1,2,3}$ also vanishes, which corresponds to $\theta_2 \rightarrow \pi/2$ in Fig.\ref{fig:6pt_diagrams}. This opening up of $\theta_2$ is a co-dimension one boundary, which corresponds to the factorization singularity of the amplitude as shown in Fig.\ref{fig:6pt_fac}. 
While $\theta_1 \rightarrow 0$ corresponds to $S_{1,2} = \la Y12 \ra^2$ vanishing. This is associated with the soft singularity, where leg-$1$ and leg-$2$ decouple (connecting with the rest of the diagram through a soft a gluon), and the remaining particles form a reducible bubble as shown in Fig.\ref{fig:6pt_soft}.
Therefore, $\theta_1 \rightarrow 0$ or $S_{1,2} \rightarrow 0$ is a co-dimension two boundary.  In a similar fashion, we find the vanishing of each $S_{i,i+1, i+2}$ leads to co-dimension one boundary, and the vanishing of $S_{i,i+1}$ corresponds to co-dimension two boundaries.

\begin{figure}
\begin{center} 
	\begin{tikzpicture}[scale=0.8, line width=1 pt]
\draw [3Dline] (1.5,1)--(0,0);
	\draw [3Dline] (0,0)--(-1.5,1);
	\draw [3Dline] (-1.5,1)--(1.5,1);
	\draw [3Dline] (0,0)--(1.5,-1);
	\draw [3Dline] (-1.5,-1)--(0,0);
	\draw [3Dline] (1.5,-1)--(-1.5,-1);
    \draw [3Dline] (-2.2,1.6)--(-1.5,1);
	\draw [3Dline] (-1.5,1)--(-2.3,0.7);
	\draw [3Dline] (1.5,1)--(2.2,1.6);
	\draw [3Dline] (2.3,0.7)--(1.5,1);
	\draw [3Dline] (-1.5,-1)--(-2.2,-1.6);
	\draw [3Dline] (-2.3,-0.7)--(-1.5,-1);
	\draw [3Dline] (2.2,-1.6)--(1.5,-1);
	\draw [3Dline] (1.5,-1)--(2.3,-0.7);
		\node at (0,0.4) {$\theta_3$};
		\node at (-2.5,1.7) {$1$};
		\node at (-2.5,0.6) {$8$};
	\end{tikzpicture}
	\begin{tikzpicture}[scale=0.8, line width=1 pt]
		\node at (0,1.5) {$\xrightarrow[]{S_{8,1,2,3}\rightarrow 0}$};
		\node at (0,0) {};
	\end{tikzpicture}
	\begin{tikzpicture}[scale=0.8, line width=1 pt]
	\draw [3Dline] (-2.2,1.6)--(-1.5,1);
	\draw [3Dline] (-1.5,1)--(-2.3,0.7);
	\draw [3Dline] (1.5,1)--(2.2,1.6);
	\draw [3Dline] (2.3,0.7)--(1.5,1);
	\draw [3Dline] (-1.5,-1)--(-2.2,-1.6);
	\draw [3Dline] (-2.3,-0.7)--(-1.5,-1);
	\draw [3Dline] (2.2,-1.6)--(1.5,-1);
	\draw [3Dline] (1.5,-1)--(2.3,-0.7);
	\path [draw=black,postaction={on each segment={mid arrow=black}}]
		
  (-1.5,1) to [bend left] (1.5,1)
   (1.5,1) to [bend left] (-1.5,1)
    (-1.5,-1) to [bend left] (1.5,-1)
     (1.5,-1) to [bend left] (-1.5,-1)
    ;
		\node at (-2.5,1.7) {$1$};
		\node at (-2.5,0.6) {$8$};
		\end{tikzpicture}  
\end{center}
\caption{The diagram shows the ${S_{8,1,2,3}\rightarrow 0}$ limit for an eight-point diagram, where the diagram separates into two reducible bubbles, therefore it is a co-dimension three boundaries.}
\label{fig:8pt_codim_3}
\end{figure}
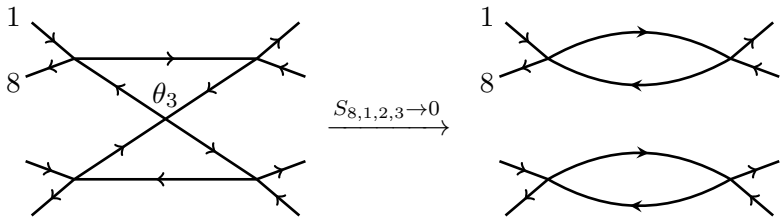

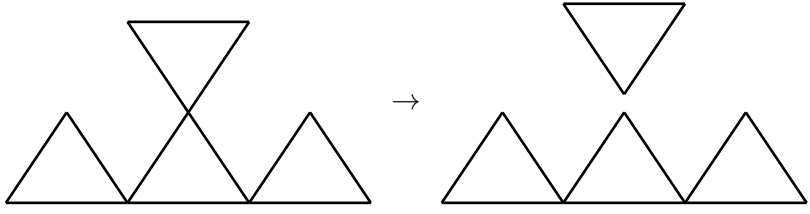
\begin{figure}
\begin{center} 
	\begin{tikzpicture}[scale=0.8, line width=1 pt]
\draw [3Dlinenoarrow] (-1,0)--(1,0);
\draw [3Dlinenoarrow] (-1,0)--(0,1.5);
\draw [3Dlinenoarrow] (1,0)--(0,1.5);
\draw [3Dlinenoarrow] (1,0)--(3,0);
\draw [3Dlinenoarrow] (1,0)--(2,1.5);
\draw [3Dlinenoarrow] (3,0)--(2,1.5);
\draw [3Dlinenoarrow] (-1,0)--(-3,0);
\draw [3Dlinenoarrow] (-1,0)--(-2,1.5);
\draw [3Dlinenoarrow] (-3,0)--(-2,1.5);

\draw [3Dlinenoarrow] (-1,3)--(1,3);
\draw [3Dlinenoarrow] (-1,3)--(0,1.5);
\draw [3Dlinenoarrow] (1,3)--(0,1.5);
		
	\end{tikzpicture}
	\begin{tikzpicture}[scale=0.8, line width=1 pt]
		\node at (0,1.5) {$\rightarrow$};
		\node at (0,0) {};
	\end{tikzpicture}
	\begin{tikzpicture}[scale=0.8, line width=1 pt]
	\draw [3Dlinenoarrow] (-1,0)--(1,0);
\draw [3Dlinenoarrow] (-1,0)--(0,1.5);
\draw [3Dlinenoarrow] (1,0)--(0,1.5);
\draw [3Dlinenoarrow] (1,0)--(3,0);
\draw [3Dlinenoarrow] (1,0)--(2,1.5);
\draw [3Dlinenoarrow] (3,0)--(2,1.5);
\draw [3Dlinenoarrow] (-1,0)--(-3,0);
\draw [3Dlinenoarrow] (-1,0)--(-2,1.5);
\draw [3Dlinenoarrow] (-3,0)--(-2,1.5);

\draw [3Dlinenoarrow] (-1,3.3)--(1,3.3);
\draw [3Dlinenoarrow] (-1,3.3)--(0,1.8);
\draw [3Dlinenoarrow] (1,3.3)--(0,1.8);
		\end{tikzpicture}  
\end{center}
\caption{The diagrams show when an even planar Mandelstam variable vanishes, the diagram on the left separates into two parts as shown on the right, and each of them contains a reducible bubble.  Note that here we omit the external legs in the diagram, which should be attached to external corners of triangles.}
\label{fig:12pt_codim_3}
\end{figure}

It is straightforward to generalize the analysis to the cases with arbitrary multiplicity. In general, we find that all odd planar Mandelstam variables correspond to co-dimension one boundaries, associated with the factorization poles of the amplitudes; whereas the two-particle planar Mandelstam variables correspond to co-dimension two boundaries, as we have seen in the six-point case. Let us now focus on the higher even planar Mandelstam variables.  This is easy to illustrate using the eight-point case as shown in Fig.\ref{fig:8pt_diagrams}.  For this particular case, $\theta_3 \rightarrow \pi/2$ exposes the singularity at $S_{8,1,2,3} \rightarrow 0$. In this limit, the diagrams separate into two parts, each containing a reducible bubble as shown in Fig.\ref{fig:8pt_codim_3}. Therefore this is a co-dimension three boundary. In Fig. \ref{fig:12pt_codim_3}, we show the same structure in an example of twelve-point diagrams.  
In general, when an even planar Mandelstam variable vanishes, the diagram separates into two parts, and each of them contains a reducible bubble, which is a co-dimension three boundary. 

\subsection{Amplitudes from the canonical form}

Following \cite{Arkani-Hamed:2013jha}, to extract the amplitude from the canonical form, we localize $Y$ as follows, 
\begin{align} \label{eq:Y*}
Y^*=\begin{pmatrix}
\mathbb{0}_{2\times k} \\
\mathbb{1}_{k\times k}
\end{pmatrix} \,,
\end{align}
and the amplitude can be obtained from the volume function $\Omega^{3d}_{n,k}(Y,\Lambda)$ by setting $Y= Y^*$ and integrating out the bosonized variables: 
\begin{align} \label{eq:amp}
\mathcal{A}_{2k}^{3d}=\delta^3(p)\int d^2\phi_1\dots d^2\phi_{k}\;\Omega^{3d}_{n,k}(Y^*,\Lambda) \,.
\end{align}
Under this projection, the $Y$-brackets become usual three-dimensional spinor-helicity brackets, namely $\langle Y^* i j\rangle = \langle i j\rangle$. Furthermore, one can perform the integration over fermionic variables $\phi$ in \eqref{eq:amp} explicitly.  In the four-point case,  we have, 
\begin{align}
\int d^2 \phi_1 d^2 \phi_2 \la 1234 \ra^2 = \delta^4(\sum_{i} q_i)  \,,  
\end{align} 
where $q_i = \lambda^{\alpha}_i \eta_i^{I}$ is the supercharge. We see that when set $Y$ to be $Y^*$, $\Omega^{3d}_{4,2}(Y, \Lambda)$ becomes
\begin{align}
\delta^4(\sum_{i} q_i)  \frac{\la 13 \ra }{\la 12 \ra \la 23 \ra} \, , 
\end{align} 
which is the four-point amplitude $\mathcal{A}_4 ( \bar{\Phi}^{\mathcal{N}=4}_1, \Phi^{\mathcal{N}=4}_2, \bar{\Phi}^{\mathcal{N}=4}_3, \Phi^{\mathcal{N}=4}_4)$. 

Similarly, we find that $\Omega^{3d}_{6,3}(Y^*, \Lambda)$ gives the six-point ABJM superamplitude. In particular, to perform the integration over the auxiliary fermionic variables $\phi$, we the following integration relation,  
\begin{align}
& \int d^2 \phi_1 d^2 \phi_2 d^2 \phi_3  \Big(\sum_{i,j=1,3,5}\la Y^{*} ij \ra\la ij246 \ra+(1,3,5)\leftrightarrow (2,4,6) \Big)^2 \nonumber\\[0.2cm]
= & \;\delta^4(\sum_{i} q_i)\, \delta^2 \Big(\sum_{i,j,k=1,3,5}\epsilon_{ijk} \la ij \ra \eta_k + (1,3,5)\leftrightarrow (2,4,6)\Big) \, .
\end{align} 
The tree-level six-point superamplitude is then given by summing over the contributions from both the positive and negative branches, which lead to
\begin{align} \label{eq:6ptsA}
&\mathcal{A}_6 ( \bar{\Phi}^{\mathcal{N}=4}_1, \Phi^{\mathcal{N}=4}_2, \dots, \bar{\Phi}^{\mathcal{N}=4}_5, \Phi^{\mathcal{N}=4}_6)\nonumber\\ =& \,  \frac{\delta^4(\sum_{i} q_i)\, \delta^2 \Big(\sum_{i,j,k=1,3,5}\epsilon_{ijk} \la ij \ra \eta_k + (1,3,5)\leftrightarrow (2,4,6)\Big)^2 {s}_{1,3,5}}{A_{52}^{*\,+}\, A_{36}^{*\,+} \, A_{14}^{*\,+}}  \nonumber\\
&+ \frac{\delta^4(\sum_{i} q_i)\, \delta^2 \Big(\sum_{i,j,k=1,3,5}\epsilon_{ijk} \la ij \ra \eta_k-(1,3,5)\leftrightarrow (2,4,6)\Big)^2 {s}_{1,3,5}}{A_{52}^{*\,-}\, A_{36}^{*\,-} \, A_{14}^{*\,-}} 
\, ,
\end{align} 
where $s_{1,3,5}=(p_1+p_3+p_5)^2$ is the standard Mandelstam variable, and  $A_{ab}^{*\, \pm}$ are defined as
\begin{align}
A_{ab}^{*\, \pm}=\sum_{i=1,3,5}\la ia\ra\la ib \ra \pm \la a{+}2\, a{-}2\ra\la b{-}2\, b{+}2 \ra \, .
\end{align} 
The expression in \eqref{eq:6ptsA} is in agreement with the known six-point superamplitude (see, e.g. \cite{Brandhuber:2012un}), after a SUSY reduction to $\mathcal{N}=4$ and translating into the normal signature. 

Finally, we remark that since the BCFW forms of the positive orthogonal Grassmannian are known to produce tree-level amplitudes in ABJM theory \cite{Huang:2013owa, Huang:2014xza}, we therefore expect that the volume forms of the orthogonal momentum amplituhedron should also lead to the correct  tree-level amplitudes for general multiplicity since they are obtained directly from the  BCFW forms.

\section{Outlook}
\label{sec:con}
In this paper, we have introduced the amplituhedron geometry associated with tree-level ABJM amplitudes. Note that through the sign flipping definition, we see that the geometry can be identified with half of the amplituhedron for four-dimensional $\mathcal{N}=4$ sYM, subject to additional momentum conservation constraint. Thus this in a sense constitutes an holographic relation, where the subspace of the four-dimension geometry lives the geometry for the three-dimensional theory. It is then natural to ask how the two forms can be related. Indeed as explored in~\cite{Huang:2014xza} the cells of orthogonal Grassmannian, and hence the associated forms, can be identified as subspace of positive cells for the usual positive Grassmannian. The form for $\mathcal{N}=4$ sYM lives in the space $Y$ and $\tilde{Y}$, it is tempting to simply identify $Y$ and $\tilde{Y}$ for $k=\frac{n}{2}$, where their dimensions are the same. However, this naive prescription cannot be the whole story since their sign flipping conditions are different. We leave the correct map between the two forms to future work. 

This suggests that  a similar projection might be applicable to the momentum twistor amplituhedron of $\mathcal{N}=4$ sYM. The momentum twistor Grassmannian for ABJM was studied in~\cite{Elvang:2014fja}, where it was found that the orthogonal condition is defined on a kinematic dependent metric. Our analysis motivates us to take the $\mathcal{N}=4$ amplituhedron, and require that the four-component variables 
\eq
z_i^A\equiv(Y^\perp\cdot Z_i)^A \, ,
\eqe
satisfy the additional Sp(4) null constraint
\eq
z_i^A z_i^B \Omega_{AB}=0\, ,
\eqe
where $\Omega_{AB}$ is the Sp(4) invariant metric. We leave the exploration of this possibility to future work as well. 

In a recent work~\cite{Song}, it was shown that the orthogonal momentum amplituhedron can be identified as the push forward of the canonical form on the moduli-space of $n$ punctured disk $\mathcal{M}^+_{0,n}$, through the Veronese map. The image has the property that the zero of even-particle Mandelstams are higher co-dimensional boundaries. However, for the pre-image, these are all co-dimension one boundaries. Thus it would appear that while the push forward maps boundary to boundary, the co-dimensionality will change. It will be interesting to understand how the Veronese map systematically achieves this and what is the  geometric mechanism behind it.

The positive orthogonal Grassmannian geometry has very intriguing connections with the correlation functions of planar Ising networks \cite{Ising1, Ising2}, and the connections have led to efficient tools for the computations of the correlation functions. It will be of interest to study if the orthogonal amplituhedron geometry constructed in this paper offers new understanding. In this paper, we extended the original amplituhedron geometry  \cite{Arkani-Hamed:2013jha}  for the scatterings in three-dimensional ABJM theory. It was understood that the scattering amplitudes in six-dimensional theories should be associated with the Symplectic Grassmannian \cite{Cachazo:2018hqa, Heydeman:2018dje, Schwarz:2019aat}, a natural future research direction is to extend the amplituhedron geometry for the Symplectic Grassmannian and study its applications for the amplitudes in six-dimensional theories.

\acknowledgments 
We would like to thank Song He, Chia-Kai Kuo and Yao-Qi Zhang for useful discussions and sharing the results of their upcoming work. Y-t Huang is supported by Taiwan Ministry of Science and Technology Grant No. 109-2112-M-002-020-MY3. R. Kojima is supported by NTU Research grant 1104000L-1. C. Wen is supported by a Royal Society University Research Fellowship No. UF160350.  S-Q Zhang is supported by the Royal Society grant RGF\textbackslash R1\textbackslash 180037.


\begin{thebibliography}{99}

\bibitem{Aharony:2008ug}
O.~Aharony, O.~Bergman, D.~L.~Jafferis and J.~Maldacena,
JHEP \textbf{10}, 091 (2008)
doi:10.1088/1126-6708/2008/10/091
[arXiv:0806.1218 [hep-th]].

\bibitem{Hosomichi:2008jb}
K.~Hosomichi, K.~M.~Lee, S.~Lee, S.~Lee and J.~Park,
JHEP \textbf{09}, 002 (2008)
doi:10.1088/1126-6708/2008/09/002
[arXiv:0806.4977 [hep-th]].


\bibitem{Witten:2003nn}
E.~Witten,
Commun. Math. Phys. \textbf{252}, 189-258 (2004)
doi:10.1007/s00220-004-1187-3
[arXiv:hep-th/0312171 [hep-th]].



\bibitem{Roiban:2004yf}
R.~Roiban, M.~Spradlin and A.~Volovich,
Phys. Rev. D \textbf{70}, 026009 (2004)
doi:10.1103/PhysRevD.70.026009
[arXiv:hep-th/0403190 [hep-th]].




\bibitem{Huang:2012vt}
Y.~t.~Huang and S.~Lee,
Phys. Rev. Lett. \textbf{109}, 191601 (2012)
doi:10.1103/PhysRevLett.109.191601
[arXiv:1207.4851 [hep-th]].



\bibitem{Drummond:2008vq} 
J.~M.~Drummond, J.~Henn, G.~P.~Korchemsky and E.~Sokatchev,
Nucl. Phys. B \textbf{828}, 317-374 (2010)
doi:10.1016/j.nuclphysb.2009.11.022
[arXiv:0807.1095 [hep-th]].

\bibitem{Drummond:2009fd}
J.~M.~Drummond, J.~M.~Henn and J.~Plefka,
JHEP \textbf{05}, 046 (2009)
doi:10.1088/1126-6708/2009/05/046
[arXiv:0902.2987 [hep-th]].


\bibitem{Bargheer:2010hn}
T.~Bargheer, F.~Loebbert and C.~Meneghelli,
Phys. Rev. D \textbf{82}, 045016 (2010)
doi:10.1103/PhysRevD.82.045016
[arXiv:1003.6120 [hep-th]].


\bibitem{Huang:2010qy}
Y.~t.~Huang and A.~E.~Lipstein,
JHEP \textbf{11}, 076 (2010)
doi:10.1007/JHEP11(2010)076
[arXiv:1008.0041 [hep-th]].

\bibitem{Gang:2010gy}
D.~Gang, Y.~t.~Huang, E.~Koh, S.~Lee and A.~E.~Lipstein,
JHEP \textbf{03}, 116 (2011)
doi:10.1007/JHEP03(2011)116
[arXiv:1012.5032 [hep-th]].

\bibitem{Arkani-Hamed:2009ljj}
N.~Arkani-Hamed, F.~Cachazo, C.~Cheung and J.~Kaplan,
JHEP \textbf{03}, 020 (2010)
doi:10.1007/JHEP03(2010)020
[arXiv:0907.5418 [hep-th]].



\bibitem{Lee:2010du}
S.~Lee,
Phys. Rev. Lett. \textbf{105}, 151603 (2010)
doi:10.1103/PhysRevLett.105.151603
[arXiv:1007.4772 [hep-th]].

\bibitem{Arkani-Hamed:2012zlh}
N.~Arkani-Hamed, J.~L.~Bourjaily, F.~Cachazo, A.~B.~Goncharov, A.~Postnikov and J.~Trnka,
doi:10.1017/CBO9781316091548
[arXiv:1212.5605 [hep-th]].






\bibitem{Huang:2013owa}
Y.~T.~Huang and C.~Wen,
JHEP \textbf{02}, 104 (2014)
doi:10.1007/JHEP02(2014)104
[arXiv:1309.3252 [hep-th]].

\bibitem{Huang:2014xza}
Y.~t.~Huang, C.~Wen and D.~Xie,
J. Phys. A \textbf{47}, no.47, 474008 (2014)
doi:10.1088/1751-8113/47/47/474008
[arXiv:1402.1479 [hep-th]].

\bibitem{Hodges:2009hk}
A.~Hodges,
JHEP \textbf{05} (2013), 135
doi:10.1007/JHEP05(2013)135
[arXiv:0905.1473 [hep-th]].


\bibitem{Arkani-Hamed:2013jha}
N.~Arkani-Hamed and J.~Trnka,
JHEP \textbf{10}, 030 (2014)
doi:10.1007/JHEP10(2014)030
[arXiv:1312.2007 [hep-th]].


\bibitem{Arkani-Hamed:2013kca}
N.~Arkani-Hamed and J.~Trnka,
JHEP \textbf{12}, 182 (2014)
doi:10.1007/JHEP12(2014)182
[arXiv:1312.7878 [hep-th]].




\bibitem{Damgaard:2019ztj} 
D.~Damgaard, L.~Ferro, T.~Lukowski and M.~Parisi,
``The Momentum Amplituhedron,''
JHEP \textbf{08}, 042 (2019)
doi:10.1007/JHEP08(2019)042
[arXiv:1905.04216 [hep-th]].



\bibitem{He:2018okq}
S.~He and C.~Zhang,
``Notes on Scattering Amplitudes as Differential Forms,''
JHEP \textbf{10}, 054 (2018)
doi:10.1007/JHEP10(2018)054
[arXiv:1807.11051 [hep-th]].

\bibitem{Ferro:2020lgp}
L.~Ferro, T.~\L{}ukowski and R.~Moerman,
JHEP \textbf{07}, no.07, 201 (2020)
doi:10.1007/JHEP07(2020)201
[arXiv:2003.13704 [hep-th]].

\bibitem{Damgaard:2020eox}
D.~Damgaard, L.~Ferro, T.~Lukowski and R.~Moerman,
JHEP \textbf{02}, 041 (2021)
doi:10.1007/JHEP02(2021)041
[arXiv:2010.15858 [hep-th]].

\bibitem{Damgaard:2021qbi}
D.~Damgaard, L.~Ferro, T.~Lukowski and R.~Moerman,
JHEP \textbf{07}, 111 (2021)
doi:10.1007/JHEP07(2021)111
[arXiv:2103.13908 [hep-th]].




\bibitem{Arkani-Hamed:2017vfh}
N.~Arkani-Hamed, H.~Thomas and J.~Trnka,
JHEP \textbf{01} (2018), 016
doi:10.1007/JHEP01(2018)016
[arXiv:1704.05069 [hep-th]].

\bibitem{Song}
S.~He, C.~K.~Kuo and Y.~Q.~Zhang,
``The momentum amplituhedron of SYM and ABJM from twistor-string maps,''
[arXiv:2111.02576 [hep-th]].

\bibitem{Brandhuber:2012un}
A.~Brandhuber, G.~Travaglini and C.~Wen,
JHEP \textbf{07}, 160 (2012)
doi:10.1007/JHEP07(2012)160
[arXiv:1205.6705 [hep-th]].


\bibitem{Elvang:2014fja}
H.~Elvang, Y.~t.~Huang, C.~Keeler, T.~Lam, T.~M.~Olson, S.~B.~Roland and D.~E.~Speyer,
JHEP \textbf{12}, 181 (2014)
doi:10.1007/JHEP12(2014)181
[arXiv:1410.0621 [hep-th]].




\bibitem{Ising1}
P. Galashin and P. Pylyavskyy 
`` Ising model and the positive orthogonal Grassmannian", arXiv:1807.03282. 

\bibitem{Ising2}
Y.~T.~Huang, C.~K.~Kuo and C.~Wen,
Phys. Rev. Lett. \textbf{121} (2018) no.25, 251604
doi:10.1103/PhysRevLett.121.251604
[arXiv:1809.01231 [hep-th]].


\bibitem{Cachazo:2018hqa}
F.~Cachazo, A.~Guevara, M.~Heydeman, S.~Mizera, J.~H.~Schwarz and C.~Wen,
JHEP \textbf{09} (2018), 125
doi:10.1007/JHEP09(2018)125
[arXiv:1805.11111 [hep-th]].

\bibitem{Heydeman:2018dje}
M.~Heydeman, J.~H.~Schwarz, C.~Wen and S.~Q.~Zhang,
Phys. Rev. Lett. \textbf{122} (2019) no.11, 111604
doi:10.1103/PhysRevLett.122.111604
[arXiv:1812.06111 [hep-th]].

\bibitem{Schwarz:2019aat}
J.~H.~Schwarz and C.~Wen,
JHEP \textbf{08} (2019), 125
doi:10.1007/JHEP08(2019)125
[arXiv:1907.03485 [hep-th]].

\end{thebibliography}
\end{document}